\documentclass[a4paper,12pt]{article}
\usepackage{amsmath}
\usepackage{amssymb}
\usepackage{amsfonts}
\usepackage{fleqn}
\usepackage[footnotesize]{caption}
\usepackage{mathrsfs}

\def\SU{\text{SU}}

\def\beq{\begin{equation}}
\def\eeq{\end{equation}}
\def\bea{\begin{eqnarray}}
\def\eea{\end{eqnarray}}

\def\<{\left\langle}
\def\>{\right\rangle}

\allowdisplaybreaks[2]

\begin{document}

\bibliographystyle{OurBibTeX}

\begin{titlepage}

 \vspace*{-15mm}
\begin{flushright}
hep-ph/0506297\\
\end{flushright}
\vspace*{5mm}

\begin{center}
{
\sffamily
\LARGE
Predicting neutrino parameters from
$SO(3)$ family symmetry and quark-lepton unification}
\\[8mm]
S.~F.~King\footnote{E-mail: \texttt{sfk@hep.phys.soton.ac.uk}}$^{(a)}$
\\[3mm]
{\small\it $^{(a)}$
School of Physics and Astronomy,
University of Southampton,\\
Southampton, SO17 1BJ, U.K.
}\\[1mm]
\end{center}
\vspace*{0.75cm}

\begin{abstract}

\noindent
We show how the neutrino mixing angles and oscillation phase
can be predicted from tri-bimaximal neutrino mixing,
corrected by charged lepton mixing angles which are related
to quark mixing angles via quark-lepton unification.
The tri-bimaximal neutrino mixing can naturally originate from the see-saw mechanism
via constrained sequential dominance (CSD), where CSD can result from
the vacuum alignment of a non-Abelian family symmetry such as $SO(3)$.
We construct a realistic model of
quark and lepton masses and mixings based on $SO(3)$ family
symmetry with quark-lepton unification based on the Pati-Salam gauge group.
The atmospheric angle is predicted to be approximately maximal
$\theta_{23}= 45^\circ$, corrected by
the quark mixing angle $\theta_{23}^{\mathrm{CKM}}\approx 2.4^\circ$,
with the correction controlled by an undetermined phase in the quark sector.
The solar angle is predicted by the tri-bimaximal complementarity
relation: $\theta_{12}+ \frac{1}{\sqrt{2}}\frac{\theta_{\mathrm{C}}}{3}
\cos (\delta - \pi) \approx 35.26^\circ $, where $\theta_{\mathrm{C}}$ is the
Cabibbo angle and $\delta$ is the neutrino oscillation phase.
The reactor angle is predicted to be
$\theta_{13} \approx
\frac{1}{\sqrt{2}}\frac{\theta_{\mathrm{C}}}{3}\approx 3.06^\circ$.
The MNS neutrino oscillation phase $\delta$ is predicted in terms of the solar angle
to be $\cos (\delta - \pi) \approx  (35.26^\circ - \theta_{12}^\circ )/3.06^\circ $.
These predictions can all be tested by future high precision
neutrino oscillation experiments, thereby probing the nature of high energy
quark-lepton unification.

\end{abstract}

\end{titlepage}
\newpage
\setcounter{footnote}{0}

\section{Introduction}
The discovery of neutrino masses and mixing angles, arguably the greatest
advance in physics over the past decade, has provided new clues in the search for
a theory of quark and lepton masses and mixings. For example, it is interesting to compare
the observed or bounded lepton mixing angles \cite{Maltoni:2004ei}:
\beq
\theta_{12}=33.2^\circ \pm 5^\circ , \ \ \theta_{23}=45^\circ \pm 10^\circ ,
 \ \ \theta_{13}< 13^\circ ,
\eeq
to the observed quark mixing angles and phase\cite{RPP}:
\beq
\theta^{\rm CKM}_{12}=13.0^\circ \pm 0.1^\circ , \ \ \theta^{\rm CKM}_{23}=2.4^\circ \pm 0.1^\circ , \ \
\theta^{\rm CKM}_{13}=0.2^\circ \pm 0.1^\circ , \ \ \delta^{\rm CKM}=60^\circ \pm 14^\circ.
\eeq
The quest to understand the relation between the very different lepton and quark
mixing angles has led to a great deal of theoretical model building \cite{King:2003jb}.
The poorly determined lepton parameters (especially the neutrino oscillation
phase $\delta$ which is completely undetermined) as compared to the
quark mixing angles, presents an opportunity to
make testable predictions in the lepton sector by relating
the lepton mixing parameters to the quark ones.
This can provide theoretical motivation for
making high precision measurements in the neutrino sector.
For instance the empirical relation between the leptonic mixing angle
$\theta_{12}$ (the solar angle) and the Cabibbo angle
$\theta_{\mathrm{C}}=\theta^{\rm CKM}_{12}$
\begin{eqnarray}\label{Eq:QLCrelation}
\theta_{12}+\theta_{\mathrm{C}} \approx \frac{\pi}{4}
\end{eqnarray}
has recently been the subject of much speculation
\cite{Minakata:2005rf,Raidal:2004iw,Minakata:2004xt,mohfram,QLCliterature,Antusch:2005ca,Lindner:2005pk}.
The interest arises from the possibility that this so-called
quark-lepton complementarity (QLC) relation could be a signal
of some high scale quark-lepton unification.
All the attempts to reproduce the QLC relation in the literature
so far start from some kind of maximal or bi-maximal mixing in either
the neutrino or the charged lepton sectors, then consider the
corrections to maximal mixing coming from the other sector.

For example, in the context of inverted hierarchy models with a
pseudo-Dirac structure, it was observed some time ago
\cite{King:2000ce,King:2002nf} that the predictions for the
neutrino mixing angles of $\theta^\nu_{12}=\pi/4$,
$\theta^\nu_{13}=0$, may receive corrections from the charged
lepton mixing angle of order the Cabibbo angle, $\theta^e_{12}\sim
\theta_{\mathrm{C}}$, resulting in $\theta_{12}$ being in the LMA
MSW range, and $\theta_{13}$ close to its current experimental
limit. Recently \cite{Antusch:2005ca} it was shown that such a
scheme, when combined with a Pati-Salam symmetry, could lead to
approximate QLC. However the way that this was achieved was quite
non-trivial. The contribution to $\theta_{12}$ coming from the
charged lepton mixing angle $\theta^e_{12}$ is suppressed by a
factor of $1/\sqrt{2}$ \cite{King:2002nf}, due to the
approximately maximal atmospheric mixing angle, and an approximate
QLC relation was achieved by selecting operators which give rise
to $\theta^e_{12}\approx (3/2)\theta_C$, enhancing the charged
lepton mixing angle by a Clebsch factor of $3/2$, in order to
approximately cancel the suppression factor of $1/\sqrt{2}$
\cite{Antusch:2005ca}. This approach leads to the predictions
$m_{\mu}/m_s=2$ at the GUT scale, and the ``reactor'' leptonic
mixing angle $\theta_{13} \approx \theta_C$, both of which are on
the edge of current experimental limits. The traditional
expectation from unified models that $\theta^e_{12}\approx
\theta_{\mathrm{C}}/3$, corresponding to the Georgi-Jarlskog (GJ)
\cite{Georgi:1979df} relation $m_{\mu}/m_s=3$ at the GUT scale,
while being more consistent with data, is clearly inconsistent
with the above approach to QLC. This motivates the search for
alternative models of QLC which would be consistent with the GJ
relations.

In this paper we discuss QLC from a completely different starting point,
namely tri-bimaximal neutrino mixing\cite{tribi}.
We emphasise that, unlike \cite{tribi},
tri-bimaximal mixing {\em in the neutrino sector} is merely a
staging point in our considerations, and the final form of the
lepton mixing matrix, after charged lepton mixing angles have been
taken into account, will not have the tri-bimaximal form.
We therefore refer to this approach as
tri-bimaximal {\em complementarity} to distinguish it from the
usual tri-bimaximal neutrino mixing. To be precise we shall derive from
the see-saw mechanism
a {\em neutrino} mixing matrix of the tri-bimaximal form:
\begin{eqnarray}
V^\dagger_{\nu_\mathrm{L}} \approx
\begin{pmatrix}
\sqrt{\frac{2}{3}}  & \frac{1}{\sqrt{3}} & 0 \\
-\frac{1}{\sqrt{6}}  & \frac{1}{\sqrt{3}} & \frac{1}{\sqrt{2}} \\
\frac{1}{\sqrt{6}}  & -\frac{1}{\sqrt{3}} & \frac{1}{\sqrt{2}} \\
\end{pmatrix}
\label{tribi}
\end{eqnarray}
Then, in the conventions of Appendix A, the MNS matrix is given by
$U_{\mathrm{MNS}} = V_{e_\mathrm{L}} V^\dagger_{\nu_\mathrm{L}}$,
and so the MNS matrix will not be of the tri-bimaximal form but
will involve a left multiplication by the charged lepton mixing
matrix $V_{e_\mathrm{L}}$. Whereas the neutrino mixing angles
arising from tri-bimaximal mixing take the approximate values
$\theta^\nu_{12}=\sin^{-1}(1/\sqrt{3})=35.26^\circ$,
$\theta^\nu_{23}=45^\circ$, $\theta^\nu_{13}=0^\circ$, the
physical lepton mixing angles arising from tri-bimaximal {\em
complementarity} will differ from these values due to the charged
lepton mixing angle corrections, which in turn are related to the
quark mixing angles.
The atmospheric angle is predicted to be approximately maximal
$\theta_{23}= 45^\circ$, corrected by
the quark mixing angle $\theta_{23}^{\mathrm{CKM}}\approx 2.4^\circ$,
with the correction controlled by an undetermined phase in the quark sector.
The reactor angle is predicted to be
$\theta_{13} \approx \frac{1}{\sqrt{2}}\frac{\theta_{\mathrm{C}}}{3}\approx 3.06^\circ$.
\footnote{This prediction for the reactor angle
also follows from bi-maximal neutrino mixing,
in which $\theta^\nu_{13}=0$, and $\theta_{13}$ originates from
charged lepton mixing angle of order one third
of the Cabibbo angle, $\theta^e_{12}\sim \theta_{\mathrm{C}}/3$,
typical of the GJ correction \cite{Ferrandis:2004mq}.
However the solar angle cannot
be accounted for by such charged lepton corrections in
bi-maximal neutrino mixing \cite{Ferrandis:2004mq}.}
The solar angle is predicted from the tri-bimaximal
complementarity relation,
\begin{eqnarray}\label{Eq:TBQLCrelation}
\theta_{12}+ \frac{1}{\sqrt{2}}\frac{\theta_{\mathrm{C}}}{3}
\cos (\delta - \pi) \approx
35.26^\circ .
\end{eqnarray}
In Eq.\ref{Eq:TBQLCrelation}
the factor of $1/3$ arises from the GJ relations,
the factor of $1/\sqrt{2}$ arises
from the atmospheric angle as discussed previously, and
$\delta$ is the MNS oscillation phase.
The tri-bimaximal complementarity relation in Eq.\ref{Eq:TBQLCrelation}
may be compared to the bimaximal complementarity relation
Eq.\ref{Eq:QLCrelation}.
The two relations are approximately numerically equivalent in the case
that $\delta = \pi $ since
$\frac{1}{\sqrt{2}}\frac{\theta_{\mathrm{C}}}{3}\approx 3.06^\circ$.
\footnote{Typically
the bi-maximal complementarity relation
in Eq.\ref{Eq:QLCrelation} will also involve
a similar phase which is often neglected without good reason.
Note that the terminology ``tri-bimaximal complementarity'' as introduced here
is a short-hand for ``charged lepton corrections to tri-bimaximal neutrino mixing
in which the charged lepton mixing angles are related to the quark mixing
angles''. In particular it refers to the relation in Eq.\ref{Eq:TBQLCrelation}.''}
The tri-bimaximal complementarity relation
has the twin advantages that it incorporates the
GJ relations, as well as the factor of $1/\sqrt{2}$ which proves troublesome
for bi-maximal complementarity. In addition Eq.\ref{Eq:TBQLCrelation}
may be used to predict the neutrino oscillation phase $\delta$
from a future accurate measurement of the solar angle $\theta_{12}$,
$\cos (\delta - \pi) \approx  (35.26^\circ - \theta_{12}^\circ )/3.06^\circ $.
These predictions can be tested by future high precision
neutrino oscillation experiments.

There has recently been some progress with achieving tri-bimaximal
neutrino mixing from the see-saw mechanism using vacuum alignment with various family
symmetries such as $SU(3)$ \cite{GG} or the discrete symmetry
$A_4$ \cite{Altarelli:2005yp}. Here we shall show how
tri-bimaximal neutrino mixing can emerge in a natural and general
way from the see-saw mechanism using sequential dominance \cite{King:1998jw},
with certain simple constraints imposed on the Yukawa couplings, independently
of any particular choice of family symmetry. We refer to this
general approach as constrained sequential dominance (CSD).
CSD can arise from the vacuum alignment some non-Abelian family symmetry,
and here we focus on $SO(3)$. A potential advantage of using $SO(3)$
family symmetry is that it is possible to ``up-grade'' any
resulting model of hierarchical neutrino masses to a type II see-saw model
with a quasi-degenerate spectrum of neutrino masses
\cite{Antusch:2004xd}, and improved prospects for
leptogenesis \cite{Antusch:2004xy}, although in this paper we shall restrict
ourselves to hierarchical neutrino masses. We shall subsequently
present an explicit model based on $SO(3)$ family symmetry and
Pati-Salam unification which is consistent with all quark and
lepton masses and mixings, and gives rise to tri-bimaximal {\em
complementarity} using the Georgi-Jarlskog relations.
All types of complementarity
are crucially dependent on the plethora of complex phases which are
generally present in the Yukawa matrices. Here we shall keep careful track
of all the phases and in our approach show how tri-bimaximal complementarity may be linked
to the MNS CP violating phase.

This paper has been organized as follows: in Sec.~2, we discuss
how to achieve tri-bimaximal mixing using CSD, and briefly show
how CSD could arise from vacuum alignment with an $SO(3)$ family
symmetry. In Sec.~3 we present an explicit model based on $SO(3)$
famly symmetry and Pati-Salam unification which has all the
necessary ingredients that we require. In Sec.~4 we discuss the
predictions arising from the model, including a
careful discussion of the complex phases which appear in
tri-bimaximal complementarity.
Sec.~5 concludes the paper. In Appendix A we specify our conventions,
while in Appendix B we discuss vacuum alignment in the model.

\section{Tri-bimaximal neutrino mixing from the see-saw mechanism with constrained
sequential dominance} \label{Sec:CSD} The fact that the
tri-bimaximal neutrino mixing matrix in Eq.\ref{tribi} involves
square roots of simple ratios motivates models in which the mixing
angles are independent of the mass eigenvalues. One such class of
models are see-saw models with sequential dominance (SD) of
right-handed neutrinos \cite{King:2002nf,King:1998jw}. In SD, the atmospheric
and solar neutrino mixing angles are determined in terms of ratios
of Yukawa couplings involving the dominant and subdominant
right-handed neutrinos, respectively. If these Yukawa couplings
are simply related in some way, then it is possible for simple
neutrino mixing angle relations, such as appear in tri-bimaximal
neutrino mixing, to emerge in a simple and natural way,
independently of the neutrino mass eigenvalues.

To see how tri-bimaximal neutrino mixing could emerge from SD, we
begin by writing the right-handed neutrino Majorana mass matrix
$M_{\mathrm{RR}}$ in a diagonal basis as
\begin{eqnarray}
M_{\mathrm{RR}}\approx\begin{pmatrix}
Y  & 0 & 0 \\
0  & X & 0 \\
0  & 0 & X' \\
\end{pmatrix},
\end{eqnarray}
where we shall assume
\begin{equation}
Y\ll X\ll X'.
\label{LSD}
\end{equation}
Then in this basis we write the neutrino (Dirac) Yukawa matrix
$Y^\nu_{LR}$ in terms of the complex Yukawa couplings $a,b,c,d,e,f,a',b',c'$ as
\begin{eqnarray}
Y^{\nu}_{\mathrm{LR}} = \begin{pmatrix}
d  & a & a' \\
e  & b & b' \\
f  & c & c' \\
\end{pmatrix}.
\label{Dirac}
\end{eqnarray}
in the convention where the Yukawa matrix corresponds to the
Lagrangian coupling $\bar{L}H_uY^{\nu}_{LR}\nu_R$, where $L$ are
the left-handed lepton doublets, $H_u$ is the Higgs doublet
coupling to up-type quarks and neutrinos, and $\nu_R$ are the
right-handed neutrinos. The Dirac neutrino mass matrix is then
given by $m^{\nu}_{\mathrm{LR}} = Y^\nu_{LR} v_\mathrm{u}$, where
$v_\mathrm{u}$ is the vacuum expectation value (VEV) of $H_u$.

For simplicity we shall henceforth assume that $d=0$, although
this is not strictly necessary \cite{King:1998jw}. Then the
condition for sequential dominance (SD) is that the right-handed
neutrino of mass $Y$ gives the dominant contribution to the
see-saw mechanism, while the right-handed neutrino of mass $X$
gives the leading sub-dominant contribution \cite{King:1998jw}
\begin{equation}
\frac{|e^2|,|f^2|,|ef|}{Y}\gg
\frac{|xy|}{X} \gg
\frac{|x'y'|}{X'}
\label{srhnd}
\end{equation}
where $x,y\in a,b,c$ and $x',y'\in a',b',c'$, and all Yukawa
couplings are assumed to be complex. The combination of
Eqs.\ref{LSD},\ref{srhnd} is called light sequential dominance
(LSD) since the lightest right-handed neutrino makes the dominant
contribution to the see-saw mechanism. LSD is motivated by unified
models in which only small mixing angles are present in the Yukawa
sector, and implies that the heaviest right-handed neutrino of
mass $X'$ is irrelevant for both leptogenesis and neutrino
oscillations (for a discussion of all these points see
\cite{King:2003jb}). In addition many realistic models in the
literature (see for example \cite{Roberts:2001zy}) involve an
approximate texture zero in the 11 position, corresponding to our
simplifying assumption $d=0$. This will have the effect of
removing one of the see-saw phases.
%In the following we shall
%assume for simplicity that all the phases arising from the charged
%lepton sector to be real, so any phases present in the MNS matrix
%originate from the neutrino sector.
%\footnote{See previous footnote.}

Assuming Eq.\ref{srhnd} the neutrino masses are given to leading
order in $m_2/m_3$ by the results in \cite{King:2002nf},
summarized as:
\begin{eqnarray}
m_1 & \sim & O(\frac{x'y'}{X'})v_u^2 \label{m1} \\
m_2 & \approx &  \frac{|a|^2}{X({s^{\nu}_{12}})^2}v_u^2  \label{m2} \\
m_3 & \approx & \frac{(|e|^2+|f|^2)}{Y}v_u^2 \label{m3}
\end{eqnarray}
where $s^{\nu}_{12}=\sin \theta^{\nu}_{12}$ may be obtained from
the further results given below. Note that with SD each neutrino mass is
generated by a separate right-handed neutrino, and the sequential
dominance condition naturally results in a neutrino mass hierarchy
$m_1\ll m_2\ll m_3$. The neutrino mixing angles are given to
leading order as \cite{King:2002nf},
\begin{eqnarray}
\tan \theta^{\nu}_{23} & \approx & \frac{|e|}{|f|} \label{23}\\
\tan \theta^{\nu}_{12} & \approx &
\frac{|a|}
{c^{\nu}_{23}|b|
\cos({\phi}'_b)-
s^{\nu}_{23}|c|
\cos({\phi}'_c)} \label{12} \\
\theta^{\nu}_{13} & \approx &
e^{i(\phi^\nu_2+\phi_a-\phi_e)}
\frac{|a|(e^*b+f^*c)}{[|e|^2+|f|^2]^{3/2}}
\frac{Y}{X}
\label{13}
\end{eqnarray}
where we have written some (but not all) complex Yukawa couplings as
$x=|x|e^{i\phi_x}$. The phase $\chi^{\nu}$
is fixed to give a real angle
$\theta^{\nu}_{12}$ by,
\begin{equation}
c^{\nu}_{23}|b|
\sin({\phi'}_b)
\approx
s^{\nu}_{23}|c|
\sin({\phi'}_c)
\label{chi1}
\end{equation}
where
\begin{eqnarray}
{\phi}'_b &\equiv &
\phi_b-\phi_a-\phi^\nu_2-\chi^{\nu}, \nonumber \\
{\phi}'_c &\equiv &
\phi_c-\phi_a+\phi_e-\phi_f-\phi^\nu_2-\chi^{\nu}.
\label{bpcp}
\end{eqnarray}
The phase $\phi^\nu_2$ is fixed to give a real angle
$\theta^{\nu}_{13}$ by \cite{King:2002nf},
\begin{equation}
\phi^\nu_2 \approx  \phi_e-\phi_a -\phi_{\rm COSMO}
\label{phi2dsmall}
\end{equation}
where
\begin{equation}
\phi_{\rm COSMO}=\arg(e^*b+f^*c)
\label{lepto0}
\end{equation}
is the leptogenesis phase
corresponding to the interference diagram involving the
lightest and next-to-lightest right-handed neutrinos
\cite{King:2002nf}. The auxiliary phases appearing above are defined in Appendix A.

We can now ask what are the conditions for achieving tri-bimaximal
neutrino mixing as in Eq.\ref{tribi}, in which $\tan \theta^{\nu}_{23}=1$,
$\tan \theta^{\nu}_{12}=1/\sqrt{2}$ and $\theta^{\nu}_{13}=0$
in the framework of sequential dominance?
Note that in sequential dominance the mixing angles are determined by
ratios of Yukawa couplings, and are independent of the neutrino masses.
We propose the following set of conditions
which are sufficient to achieve tri-bimaximal mixing within the
framework of sequential dominance:
\begin{subequations}\label{tribiconds}
\begin{eqnarray}
|a| &=& |b|=|c|     \label{tribicondsd}                \;, \\
|d| & = & 0         \label{tribicondse}         \; ,\\
|e| &=& |f|         \label{tribicondsa}         \; ,\\
\phi'_b &=& 0 \label{tribicondsb}                    \; ,\\
\phi'_c &=& \pi \label{tribicondsc}                    \; .
\end{eqnarray}
\nonumber
\end{subequations}
Eqs.\ref{tribicondsd}, \ref{tribicondse}, \ref{tribicondsa}
are conditions on the magnitudes of the Yukawa couplings,
while Eqs.\ref{tribicondsb}, \ref{tribicondsc} are generic
phase conditions which can be satisfied by several different
types of phase structure in the Yukawa matrix.
The condition in Eq.\ref{tribicondsa} clearly gives rise to $\tan
\theta^{\nu}_{23}=1$, as can be seen from Eq.\ref{23}. The
remaining conditions in Eq.\ref{tribiconds} result in $\tan
\theta^{\nu}_{12}=1/\sqrt{2}$ as can be seen from Eq.\ref{12}.
Eqs.\ref{tribicondsb} and \ref{tribicondsc}, together with the
definitions in Eq.\ref{bpcp}, imply the condition on the phases of
the Yukawa couplings:
\begin{equation}
\phi_c-\phi_b+\phi_e-\phi_f=\pi  \; .
\label{pi}
\end{equation}
Eq.\ref{pi}, together with
Eqs.\ref{tribicondsa},\ref{tribicondsd}, then implies that:
\begin{equation}
e^*b+f^*c=0                   \; .
\label{zero}
\end{equation}
Eq.\ref{zero} implies from Eq.\ref{13} that $\theta^{\nu}_{13}=0$.
It also implies that leptogenesis is zero at leading order, independently
of the choice of charged lepton basis \cite{King:2002nf}.
We conclude that the conditions in Eq.\ref{tribiconds}, together
with the conditions for sequential dominance, are sufficient to
result in tri-bimaximal neutrino mixing as in Eq.\ref{tribi}.
We shall refer to this as constrained sequential dominance (CSD).
Note that, with the conditions in Eq.\ref{tribiconds} satisfied, the
angle $\theta^{\nu}_{12}$ is automaticaly real, so the phase
$\chi^{\nu}$ is undetermined, and will be expected to play no part
in physics. The phase $\phi^\nu_2$ is similarly
undetermined and unphysical, since $\theta^{\nu}_{13}=0$.

Since there are undetermined phases above, it is instructive to consider
tri-bimaximal mixing as a limiting case of a known example where the phases
are determined. The example will also serve as an introduction to the
model of quark and lepton masses and mixings
discussed in the next section based on $SO(3)$ family
symmetry and Pati-Salam unification, in which a neutrino Yukawa
matrix which satisfies the conditions of CSD,
will arise. It should be emphasised that
other examples based on $SU(3)$ or discrete family symmetries may
also give rise to CSD, the general conditions for which are
given in Eqs.\ref{tribiconds},\ref{pi}.

Consider a supersymmetric theory in which the
lepton doublets $L$ are triplets of an $SO(3)$ family symmetry,
but the (CP conjugates of) right-handed neutrinos $\nu^c_i$ and Higgs doublets $H_u$
are singlets under the family symmetry \cite{Antusch:2004xd}.
Yukawa couplings arise from the superpotential terms of the form:
\beq
|y_1|e^{i\delta_1}LH_u\nu^c_1\frac{\phi_{23}}{M}
+|y_2|e^{i\delta_2}LH_u\nu^c_2\frac{\phi_{123}}{M}
+|y_3|e^{i\delta_3}LH_u\nu^c_3\frac{\phi_{3}}{M}
\label{Yuk}
\eeq
where $\phi_{23},\phi_{123},\phi_{3}$ are $SO(3)$ triplet flavon fields whose
vacuum expectation values (VEVs) break the $SO(3)$ family
symmetry, and allow Dirac neutrino mass terms to be generated.
We have written the Yukawa couplings in terms of magnitudes and
phases $|y_i|$ and $e^{i\delta_i}$, and $M$ is a real positive mass scale.
Each term in
Eq.\ref{Yuk} only involves a particular flavon superfield coupling
together with a particular right-handed neutrino superfield. This
may readily be enforced by symmetries, as we shall discuss later
in the framework of the Pati-Salam theory.

In \cite{Antusch:2004xd,Barbieri:1999km} it was also shown how to generate real flavon VEVs:
%may be generated from the superpotential terms:
%\beq
%A(\phi_{123}^2-\Lambda_1^2)+B(\phi_{23}^2-\Lambda_2^2)
%+C(\phi_{3}^2-\Lambda_3^2) \label{ABC}
%\eeq
%where $A,B,C$ are gauge singlet superfields and $\Lambda_i$ are
%independent heavy mass terms, together with positive soft mass
%squareds for the flavon fields. This results in real VEVs of the
\beq \label{abc}
\begin{array}{l}
\frac{|y_2|\phi_{123}}{M}=
\begin{pmatrix}
a \\
b \\
c \\
\end{pmatrix},  \ \ \;
\frac{|y_1|\phi_{23}}{M}=
\begin{pmatrix}
0 \\
e \\
f \\
\end{pmatrix},  \ \ \;
\frac{|y_3|\phi_{3}}{M}=
\begin{pmatrix}
0 \\
0 \\
c' \\
\end{pmatrix}  \;.
\end{array}
\eeq
When these VEVs are inserted into the couplings in
Eq.\ref{Yuk} this results in a neutrino Yukawa matrix:
\begin{eqnarray}
Y^{\nu}_{\mathrm{LR}} =
\begin{pmatrix}
0  & ae^{i\delta_2} & 0 \\
ee^{i\delta_1}  & be^{i\delta_2} & 0 \\
fe^{i\delta_1}  & ce^{i\delta_2} & c'e^{i\delta_{3}} \\
\end{pmatrix},
\label{Dirac2}
\end{eqnarray}
where here $a,b,c,e,f,c'$ are real (positive or negative) numbers.
In order to satisfy the CSD constraints in Eqs.\ref{tribiconds},\ref{pi}
it is sufficient to
show that it is possible to arrange for the real VEVs in
Eq.\ref{abc} to be aligned such that:
\begin{subequations}\label{tribiconds3}
\begin{eqnarray}
e &=& -f         \label{vaca}         \; ,\\
a &=& b =c     \label{vacb}                \; .
\end{eqnarray}
\nonumber
\end{subequations}
The phases required to ensure positive neutrino mixing angles
are then given in the limit of such a vacuum alignment by:
\footnote{
The undetermined phases $\phi^\nu_2,\chi^{\nu}$ are specified above by
slightly relaxing the conditions in Eqs.\ref{vaca},\ref{vacb},
while keeping $|b|>|c|$. However these phases are unphysical in the
case of tri-bimaximal neutrino mixing and they could equally well be set to zero.
}
\begin{subequations}\label{tribiphases}
\begin{eqnarray}
{\phi}'_b &=&
-\phi^\nu_2-\chi^{\nu}=0, \nonumber \\
{\phi}'_c &= &
\pi-\phi^\nu_2-\chi^{\nu}=\pi
\label{bpcp2}
\;, \\
\phi^\nu_2 &=& 2(\delta_1-\delta_2)
\label{phi2dsmall2}
\;, \\
\phi^\nu_3 &=& \phi^\nu_2 + \pi
\label{phi2dsmall23}
\;, \\
\omega^\nu_1 &=& \delta_3
\label{omeganu1}
\;,\\
\omega^\nu_2 &=& \delta_2
\label{omeganu2}
\;,\\
\omega^\nu_3 &=& \delta_1
\label{omeganu3}
\;,
\end{eqnarray}
\nonumber
\end{subequations}
where the phases are defined in Appendix A.
Such a vacuum alignment would then satisfy the constraints
in Eqs.\ref{tribiconds},\ref{pi}.
In order to arrange for the VEVs in Eqs.\ref{vaca},\ref{vacb}, we
need to introduce additional flavon superfields and superpotential
terms involving these and other superfields, as discussed in Appendix B.
It is clear that
$SO(3)$ family symmetry and vacuum alignment can provide
a realization of CSD and hence tri-bimaximal {\em neutrino} mixing.
To obtain tri-bimaximal {\em complementarity} we also require
quark-lepton unification which incorporates the GJ relations, and
we now turn to the construction of such a model.

\section{$SO(3)$ family symmetry and Pati-Salam unification}\label{Sec:Models}
We now construct a realistic model
based on a family symmetry $SO(3)$ and Pati-Salam unification.
The model will incorporate both the vacuum alignment in $SO(3)$
necessary to achieve tri-bimaximal neutrino mixing via constrained
sequential dominance, and also will involve the GJ relations
necessary to relate the charged lepton mixing
angle to the Cabibbo angle.
The explicit model will demonstrate that all these features
can be achieved together within a single consistent framework,
and will lead to further relations between the lepton and quark mixing
angles.

The model is a supersymmetric theory based on the
family symmetry $SO(3)$ together with the Pati-Salam gauge group \cite{ps}
\beq
G_{PS} = \SU (4)_\mathrm{PS}\times \SU (2)_\mathrm{L}\times
\SU (2)_\mathrm{R}.
\eeq
Assuming
the Pati-Salam symmetry to start with has the advantage that it explicitly
exhibits $SU(4)_{PS}$ quark-lepton and $SU(2)_{R}$ isospin symmetry,
allowing Georgi-Jarlskog factors to be generated and isospin breaking to be
controlled, while avoiding the Higgs doublet-triplet splitting problem \cite{King:1994fv}.
Quarks and leptons are unified in the $ \SU (4)_\mathrm{PS}$-quartets
$F_{i}$ and
$F^c_i$ of
$ \SU (4)_\mathrm{C}$, which are doublets of
$\SU (2)_\mathrm{L}$ and $\SU (2)_\mathrm{R}$,
respectively,
\begin{eqnarray}\label{eq:FermionsInPati-SalamModel}
F_i =
\left( \begin{array}{cccc}
u_i & u_i & u_i& \nu_i \\
d_i & d_i & d_i& e_i
\end{array} \right)
, \quad
F_j^c =
\left( \begin{array}{cccc}
u_j^c & u_j^c & u_j^c & \nu_j^c \\
d_j^c & d_j^c & d_j^c & e_j^c
\end{array} \right),
\end{eqnarray}
where $i$ and $j$ are family indices.
In addition the left-handed quarks and leptons are assigned to be triplets,
while the CP-conjugates of the right-handed quarks and leptons
are singlets under an $SO(3)$ family symmetry,
\beq
F_i\sim {\bf 3}, \ \ \ \
F_j^c\sim {\bf 1}.
\eeq
This implies in particular that the right-handed neutrinos
$\nu_j^c\in F_j^c$ are singlets under $SO(3)$,
and the lepton electroweak doublets $L_i\in F_i$ are triplets under $SO(3)$,
as assumed previously. The usual SUSY Higgs doublets $H_u,H_d$ are contained in
a single PS bi-doublet $h$, and further heavy Higgs superfields
$H, \overline{H}$ are introduced to break the Pati-Salam symmetry group
down to the Standard Model \cite{King:1997ia}.
As in the $SU(3)$ model in \cite{GG}, we include
an adjoint $\Sigma$ field which develops vevs in the $SU(4)_{PS}\times SU(2)_R$
direction which preserves the hypercharge generator $Y=T_{3R}+(B-L)/2$.
This implies that any coupling of the $\Sigma$ to a fermion and a messenger such
as $\Sigma^{a \alpha}_{b \beta}F^c_{a\alpha}\chi^{b\beta}$, where the
$SU(2)_R$ and $SU(4)_{PS}$ indices have been displayed explicitly, is
proportional to the hypercharge $Y$ of the particular fermion component of
$F^c$ times the vev $\sigma$. For example the coupling of $\Sigma$
to right-handed neutrinos gives zero.
In addition to $SO(3)\times G_{PS}$, the flavour symmetry group also
includes $R\times Z_4^2 \times Z_3^2 \times U(1)$ symmetries in order to
restrict the form of the mass matrices, where
the continuous R-symmetry may be alternatively be replaced
by a discrete $Z_{2R}$ symmetry.
The superfields transform under the full
symmetry group of the model as shown in Table \ref{Table1}.

\bigskip {\small
%TCIMACRO{\TeXButton{B}{\begin{table}[tbp] \centering}}%
%BeginExpansion
\begin{table}[tbp] \centering%
%EndExpansion
\begin{tabular}{|ccccccrrrrr|}
\hline
${\bf Field}$ & ${\bf SO(3)}$ & ${\bf SU(4)_{PS}}$ & ${\bf SU(2)_{L}}$ & $%
{\bf SU(2)_{R}}$ & ${\bf R}$ & ${\bf U(1)}$ & ${\bf Z_4^{I}}$
& ${\bf Z_4^{II}}$ & ${\bf Z_3^{I}}$ & ${\bf Z_3^{II}}$\\ \hline
$F_i$ & ${\bf 3}$ & ${\bf 4}$ & ${\bf 2}$ & ${\bf 1}$ & ${\bf 1}$ &
${\bf 0}$ & ${\bf 0}$ & ${\bf 0}$ & ${\bf 0}$ & ${\bf 0}$\\
$F_1^{c}$ & ${\bf 1}$ & ${\bf \overline{4}}$ & ${\bf 1}$ & ${\bf 2}$ &
${\bf 1}$ & ${\bf -3}$ & ${\bf \alpha}$ & ${\bf 0}$ & ${\bf 0}$ & ${\bf 0}$\\
$F_2^{c}$ & ${\bf 1}$ & ${\bf \overline{4}}$ & ${\bf 1}$ & ${\bf 2}$ &
${\bf 1}$ & ${\bf -3}$ & ${\bf 0}$ & ${\bf \beta}$ & ${\bf 0}$ & ${\bf 0}$\\
$F_3^{c}$ & ${\bf 1}$ & ${\bf \overline{4}}$ & ${\bf 1}$ & ${\bf 2}$ &
${\bf 1}$ & ${\bf 0}$ & ${\bf 0}$ & ${\bf 0}$ & ${\bf 0}$ & ${\bf 0}$\\
$h$ & ${\bf 1}$ & ${\bf 1}$ & ${\bf 2}$ & ${\bf 2}$ & ${\bf 0}$ &
${\bf 0}$ & ${\bf 0}$ & ${\bf 0}$ & ${\bf 0}$ & ${\bf 0}$\\
$H$ & ${\bf {1}}$ & ${\bf 4}$ & ${\bf 1}$ & ${\bf 2}$ & ${\bf 0}$ &
${\bf 0}$ & ${\bf 0}$ & ${\bf 0}$ & ${\bf 0}$ & ${\bf 0}$\\
$\overline{H}$ & ${\bf 1}$ & ${\bf \overline{4}}$ & ${\bf 1}$ &
${\bf 2}$ & ${\bf 0}$ & ${\bf 0}$ & ${\bf 0}$ & ${\bf 0}$ & ${\bf 0}$ & ${\bf 0}$ \\
$\Sigma $ & ${\bf 1}$ & ${\bf 15}$ & ${\bf 1}$ & ${\bf 3}$ & ${\bf 0}$ &
${\bf 2}$ & ${\bf \alpha^3}$ & ${\bf \beta^3}$ & ${\bf 0}$ & ${\bf 0}$\\
$\phi _{1}$ & ${\bf {3}}$ & ${\bf 1}$ & ${\bf 1}$ & ${\bf 1}
$ & ${\bf 0}$ & ${\bf 0}$ & ${\bf 0}$ & ${\bf 0}$ & ${\bf \gamma }$ & ${\bf 0}$\\
$\phi _{2}$ & ${\bf {3}}$ & ${\bf 1}$ & ${\bf 1}$ & ${\bf 1}
$ & ${\bf 0}$ & ${\bf 0}$ & ${\bf 0}$ & ${\bf 0}$ & ${\bf 0}$ & ${\bf \delta}$\\
$\phi _{3}$ & ${\bf {3}}$ & ${\bf 1}$ & ${\bf 1}$ & ${\bf 3\oplus 1}
$ & ${\bf 0}$ & ${\bf 0}$ & ${\bf 0}$ & ${\bf 0}$ & ${\bf 0}$ & ${\bf 0}$\\
$\phi _{23}$ & ${\bf {3}}$ & ${\bf 1}$ & ${\bf 1}$ & ${\bf 1}$ &
${\bf 0}$ & ${\bf 1}$ & ${\bf \alpha}$ & ${\bf 0}$ & ${\bf 0}$ & ${\bf 0}$\\
$\phi _{123}$ & ${\bf {3}}$ & ${\bf 1}$ & ${\bf 1}$ & ${\bf 1}$ &
${\bf 0}$ & ${\bf 1}$ & ${\bf 0}$ & ${\bf \beta}$ & ${\bf 0}$ & ${\bf 0}$\\
\hline
\end{tabular}%
\caption{{\footnotesize Transformation of the superfields under the
$SO(3)$ family, Pati-Salam and $R\times Z_4^2 \times Z_3^2 \times U(1)$ symmetries which
restrict the form of the mass matrices.
The continuous R-symmetry may be alternatively be replaced
by a discrete $Z_{2R}$ symmetry.
We only display the
fields relevant for generating fermion mass and spontaneous symmetry
breaking.  \label{Table1}}}
%TCIMACRO{\TeXButton{E}{\end{table}}}%
%BeginExpansion
\end{table}%
%EndExpansion
}

We need spontaneous breaking of the family
symmetry
\begin{equation}
SO(3)\longrightarrow SO(2)\longrightarrow {\rm Nothing}  \label{fsb}
\end{equation}%
To achieve this symmetry breaking we introduce additional flavon
fields $\phi_{i}$, $\phi _{23}, \phi _{123}$ in the
representations given in Table \ref{Table1}. The vacuum alignment
of the flavon VEVs plays a crucial role in this model. In the
$SO(3)$ model the flavon VEVs are all real and, as discussed in
Appendix B, may be aligned in the following way:
\beq \label{123}
\begin{array}{l}
\phi_{1}\sim
\begin{pmatrix}
1 \\
0 \\
0 \\
\end{pmatrix}, \ \ \;
\phi_{2}\sim
\begin{pmatrix}
0 \\
1 \\
0 \\
\end{pmatrix}, \ \ \;
\phi_{3}\sim
\begin{pmatrix}
0 \\
0 \\
1 \\
\end{pmatrix}, \ \ \;
\phi_{23}\sim
\begin{pmatrix}
0 \\
1 \\
-1 \\
\end{pmatrix},  \ \ \;
\phi_{123}\sim
\begin{pmatrix}
1 \\
1 \\
1 \\
\end{pmatrix}.
\end{array}
\eeq

The leading operators allowed by the symmetries are
\begin{eqnarray}
W_{{\rm Yuk}} &= &
\frac{y_1}{M^{3}}(F.\phi _{23})\phi_{23}^2F _{1}^{c}h
\label{op1} \\
&+&\frac{y_2}{M^3}(F.\phi _{123})\phi_{123}^2F _{2}^{c}h
+\frac{y_2'\Sigma }{M^2}(F.\phi _{23})F _{2}^{c}h
\label{op2} \\
&+&\frac{y_3}{M}(F.\phi _{3})F _{3}^{c}h
+\frac{y_3'}{M^3}(F.\phi _{2})\phi_2^2F _{3}^{c}h
+\frac{y_3''}{M^3}(F.\phi _{1})\phi_1^2F _{3}^{c}h
  \label{op3} \\
W_{{\rm Maj}} &\sim &\frac{1}{M}(F _{3}^{c} H)^2 \label{mop1} \\
&+&\frac{1}{M^7}(F _{2}^{c} H)^2(\phi_{123}^2\phi_{23}^4+\phi_{123}^6)
\label{mop2} \\
&+&\frac{1}{M^7}(F _{1}^{c} H)^2(\phi_{23}^6+\phi_{23}^2\phi_{123}^4)
\label{mop3} \\
&+&\frac{1}{M^5}(F _{2}^{c} H)(F _{3}^{c} H)\phi_{123}^3\phi_3
\label{mop4} \\
&+&\frac{1}{M^5}(F _{1}^{c} H)(F _{3}^{c} H)\phi_{23}^3\phi_3
\label{mop5} \\
&+&\frac{1}{M^7}(F _{1}^{c} H)(F _{2}^{c} H)\phi_{23}^3\phi_{123}^3,
\label{mop6}
\end{eqnarray}
where we have included complex Yukawa couplings
$y_i=|y_i|e^{i\delta_i}$ in the Yukawa superpotential,
but have suppressed similar Yukawa couplings
which would appear multiplying the Majorana operators.

In order to obtain the Yukawa matrices from $W_{{\rm Yuk}}$
and the Majorana matrix from $W_{{\rm Maj}}$ requires some discussion
of the messenger sector that is responsible for the operators above.
This was fully discussed in \cite{GG}, and we shall only briefly repeat the
essential points here. The operators
arise from Froggat-Nielsen diagrams and the scale $M$
represents the right-handed up and down messenger mass scales $M^{u,d}$,
corresponding to the dominance of right-handed messengers
over left-handed messengers, which applies if $M<M^{L}$
where $M^{L}$ represents the left-handed messenger mass scale.
Specifically it was assumed that
\begin{equation}
M^{d}\approx \frac{1}{3}M^{u}\ll M^{L}.  \label{mess}
\end{equation}%
%Such a universal structure is to be expected in theories with Wilson line
%breaking in which the breaking is due to the $(4D)$ scalar component of a
%higher dimension gauge field because it couples universally to fields in the
%same representation of gauge group factors left unbroken by the Wilson line \cite{GG}.
%The Wilson line breaking is associated with the compactification and so the
%splitting induced is naturally of order the compactification scale. Thus, if
%Wilson line breaking is responsible for breaking $SU(2)_{R},$ the messenger
%states (Kaluza-Klein modes or vectorlike states obtaining mass on
%compactification) must have masses of order the compactification scale.
It was further assumed that
%$SU(4)_{PS}$ remains after compactification so that its subsequent
%breaking will be a small effect and so
the right-handed lepton messenger scales satisfy the approximate $SU(4)_{PS}$
relations $M^{\nu }\simeq M^{u}$, and $M^{e}\simeq M^{d}$. The splitting of
the messenger mass scales relies on left-right and $SU(2)_{R}$ breaking
effects which was assumed to be due to a Wilson line symmetry breaking
mechanism \cite{GG}. Eq.\ref{mess} then allows the expansion parameters associated with
$\phi_{23}$ to take the numerical values \cite{GG}:
\begin{equation}
\epsilon \equiv \frac{\phi_{23}}{M^{u}}\approx 0.05,
\ \ \bar{\epsilon}\equiv \frac{\phi_{23}}{M^{d}}\approx 0.15
\label{eps}
\end{equation}
where here and henceforth we assume that the fields have been
replaced by their VEVs e.g. $\phi_{23}\rightarrow <\phi_{23}>$, etc.
We shall also assume that the flavons $\phi_{123}$ take similar VEVs:
\begin{equation}
\frac{\phi_{123}}{M^{u}}\approx \epsilon,
\ \ \frac{\phi_{123}}{M^{d}}\approx \bar{\epsilon}
\label{eps123}
\end{equation}
In the present model the flavons $\phi _{1,2}$ lead to independent
expansion parameters which we will assume to satisfy:
\begin{equation}
\frac{\phi_{1}}{M^{d}}\approx \bar{\epsilon}, \ \
\frac{\phi_{2}}{M^{d}}\approx \bar{\epsilon}^{2/3}.
\label{a12d}
\end{equation}
Eqs.\ref{eps}, \ref{a12d} then imply:
\begin{equation}
\frac{\phi_{1}}{M^{u}}\approx {\epsilon}, \ \
\frac{\phi_{2}}{M^{u}}\approx \frac{\bar{\epsilon}^{2/3}}{3} \approx 0.7{\epsilon}^{2/3}.
\label{a12u}
\end{equation}
The flavon $\phi _{3}$ transforms under
$SU(2)_{R}$ as ${\bf 3\oplus 1}$, and develops
isospin breaking vevs in the up and down $SU(2)_{R}$ directions,
and we assume as in \cite{GG}:
\begin{equation}
\frac{\phi_{3}^{u}}{M^{u}}=\frac{\phi_{3}^{d}}{M^{d}}\approx \sqrt{\bar{\epsilon}}.
\label{a3}
\end{equation}
It remains to specify the expansion parameter associated with $\sigma $, the
vev of $\Sigma $. This was determined purely by
phenomenological considerations in \cite{GG}, and here we assume the same value:
\begin{equation}
Y(d)\frac{\sigma }{M^{d}}\approx \bar{\epsilon}.
%Y(u)\frac{\sigma }{M^{u}}\approx {\epsilon}.
 \label{sigma}
\end{equation}%
Note that the operators involving $\Sigma$
must be multiplied by the hypercharge of the relevant
right-handed fermion, where
$Y(d)=1/3$ is the hypercharge of $d^{c}$,
$Y(u)=-2/3$ is the hypercharge of $u^{c}$,
$Y(e)=1$ is the hypercharge of $e^{c}$, and
$Y(\nu)=0$ is the hypercharge of $\nu^{c}$.

The operators in Eqs.\ref{op1},\ref{op2},\ref{op3} when combined with
the messenger sector just described, then leads to the Yukawa matrices:
\begin{eqnarray}
Y_{LR}^{U} &\approx &\left(
\begin{array}{clr}
0 & y_2\epsilon ^{3} & y_3''\epsilon ^{3} \\
y_1\epsilon ^{3} & y_2\epsilon^3-{2}y_2'\epsilon ^{2}
& 0.34y_3'\epsilon ^{2}\\
-y_1\epsilon ^{3}& y_2\epsilon^3+{2}y_2'\epsilon ^{2}
& y_3\bar{\epsilon}^{\frac{1}{2}}
\end{array}\right) ,  \label{Yu} \\
Y_{LR}^{D} &\approx &\left(
\begin{array}{clr}
0 & y_2\bar{\epsilon} ^{3} & y_3''\bar{\epsilon} ^{3} \\
y_1\bar{\epsilon} ^{3} & y_2\bar{\epsilon}^3+y_2'\bar{\epsilon} ^{2}
& y_3'\bar{\epsilon} ^{2}\\
-y_1\bar{\epsilon} ^{3}& y_2\bar{\epsilon}^3-y_2'\bar{\epsilon} ^{2}
& y_3{\bar{\epsilon}}^{\frac{1}{2}}
\end{array}\right) ,  \label{Yd} \\
Y_{LR}^{E} &\approx &\left(
\begin{array}{clr}
0 & y_2\bar{\epsilon} ^{3} & y_3''\bar{\epsilon} ^{3} \\
y_1\bar{\epsilon} ^{3} & y_2\bar{\epsilon}^3+3y_2'\bar{\epsilon} ^{2}
& y_3'\bar{\epsilon} ^{2}\\
-y_1\bar{\epsilon} ^{3}& y_2\bar{\epsilon}^3-3y_2'\bar{\epsilon} ^{2}
& y_3{\bar{\epsilon}}^{\frac{1}{2}}
\end{array}\right) ,  \label{Ye} \\
Y_{LR}^{\nu} &\approx &\left(
\begin{array}{clr}
0 & y_2\epsilon ^{3} & y_3''\epsilon ^{3} \\
y_1\epsilon ^{3} & y_2\epsilon^3
& 0.34y_3'\epsilon ^{2}\\
-y_1\epsilon ^{3}& y_2\epsilon^3
& y_3\bar{\epsilon}^{\frac{1}{2}}
\end{array}\right) . \label{Ynu}
\end{eqnarray}
The leading corrections are given
by additional operators similar to those displayed but with insertions
of powers of $\phi_3^2/M^2\sim \bar{\epsilon}$.

The leading heavy right-handed neutrino Majorana mass arises from the
operator of Eq.\ref{mop1} which gives,
\begin{equation}
M_{3}\approx \frac{<H>^{2}}{M^{\nu }},  \label{M3}
\end{equation}%
to the third family, where $M^{\nu }=M^{u}$ is the same messenger mass scale
as in the up sector due to $SU(4)_{PS}$. Operators involving $\Sigma $ do
not contribute since it does not couple to right-handed neutrinos which have
zero hypercharge. However the other Majorana operators fill
out the Majorana mass matrix, and after
small angle right-handed rotations the Majorana matrix takes the form:
\begin{equation}
M_{RR}= \left(
\begin{array}{ccr}
p\epsilon ^{6} & 0  & 0 \\
0 & q \epsilon ^{6} & 0 \\
0 & 0 & 1
\end{array}
\right) M_{3},  \label{MRRA}
\end{equation}
where $p,q$ are complex couplings.

\section{Predictions for Neutrino Parameters}
\label{Sec:TextureExample}

The neutrino Yukawa matrix in Eq.\ref{Ynu} has the CSD form considered
in Eqs.~\ref{Dirac2},\ref{tribiconds3} and,
provided the SD conditions in Eq.\ref{srhnd} are satisfied,
it will lead to tri-bimaximal neutrino mixing.
The complex phases in $M_{RR}$ in Eq.\ref{MRRA} may be removed by
rotations on the right-handed neutrino fields, which only results in
a redefinition of the Yukawa phases appearing in the complex Yukawa
couplings $y_i=|y_i|e^{i\delta_i}$. Effectively, then, the Majorana masses in
$M_{RR}$ may be taken to be real without loss of generality.
The SD conditions in Eq.\ref{srhnd} are then satisfied providing:
\begin{equation}
\frac{|y_1^2|}{p}\gg
\frac{|y_2^2|}{q} \gg
|y_3^2|\bar{\epsilon}.
\label{srhnd2}
\end{equation}
Since the Yukawa couplings are expected to be of order unity, the model
predicts a rather mild hierarchy in physical neutrino masses, from Eqs.\ref{m1},
\ref{m2},\ref{m3}:
\begin{eqnarray}
m_1 & \approx & \frac{|y_3^2|\bar{\epsilon}}{M_3}v_u^2 \label{m1p} \\
m_2 & \approx &  \frac{3|y_2^2|}{qM_3}v_u^2  \label{m2p} \\
m_3 & \approx & \frac{2|y_1^2|}{pM_3}v_u^2 \label{m3p}
\end{eqnarray}
The neutrino mixing angles take the tri-bimaximal values:
\begin{eqnarray}
\tan \theta^{\nu}_{23} & \approx & 1\label{23p}\\
\tan \theta^{\nu}_{12} & \approx & \frac{1}{\sqrt{2}}
\label{12p} \\
\theta^{\nu}_{13} & \approx & 0
\label{13p}
\end{eqnarray}
since the Yukawa matrix in Eq.\ref{Ynu} has the CSD form considered
in Eqs.~\ref{Dirac2},\ref{tribiconds3}, as already discussed.
The neutrino auxiliary phases then take the values given in Eq.\ref{tribiphases},
where the phases $\delta_i$ refer to the phases of  the Yukawa couplings
$y_i=|y_i|e^{i\delta_i}$ (assuming without loss of generality real $p,q$).
From Eq.\ref{tribiphases} and Eqs.~\ref{B5}-\ref{B7} we obtain the neutrino phases:
\bea
\delta_{12}^{\nu_L}&=&(\delta_3 - \delta_2)
\label{B5p}\\
\delta_{13}^{\nu_L}&=&(\delta_3 - \delta_1)
\label{B6p}\\
\delta_{23}^{\nu_L}&=&-3(\delta_1 - \delta_2).
\label{B7p}
\eea

The lepton mixing angles receive corrections from the charged lepton sector
which in this model are completely derivable from the charged lepton Yukawa
matrix in Eq.\ref{Ye}, using the results in Appendix A.
The charged lepton Yukawa matrix in Eq.\ref{Ye} leads to charged lepton
masses in the ratios:
\bea
m_e:m_\mu :m_\tau & \approx &
\frac{|y_1||y_2|}{3|y'_2|}\bar{\epsilon}^4: 3|y'_2|\bar{\epsilon}^2:y_{33}
\label{chlepmasses}
\eea
In a small charged lepton angle approximation,
\begin{eqnarray}
\theta^{E_L}_{23} & \approx & \frac{|y'_3|\bar{\epsilon}^2}{y_{33}}
\label{23E}\\
\theta^{E_L}_{13} & \approx & \frac{|y''_3|\bar{\epsilon}^3}{y_{33}}
\label{13E} \\
\theta^{E_L}_{12} & \approx & \frac{|y_2|\bar{\epsilon}}{3|y'_2|}
\label{12E}
\end{eqnarray}
where we have written $y_{33}=|y_3|\bar{\epsilon}^{\frac{1}{2}}$.
The auxiliary charged lepton phases, used to make the charged lepton mixing angles
real and positive, are:
\begin{subequations}\label{tribiphasesE}
\begin{eqnarray}
\phi^{E_L}_2 &=& \delta'_3-\delta''_3
\label{phi2EL}
\;, \\
\phi^{E_L}_3 &=& \delta_3-\delta''_3
\label{phi3EL}
\;, \\
\chi^{E_L} &=& \delta'_2-\delta_2-\delta'_3+\delta''_3
\label{chiEL}
\;
\end{eqnarray}
\nonumber
\end{subequations}
$\omega^{E_L}_i$ are undetermined and are used to remove phases from the MNS matrix.
From Eq.\ref{tribiphases},\ref{tribiphasesE} and Eqs.~\ref{B8}-\ref{B10}
we obtain the charged lepton phases:
\bea
\delta_{23}^{E_L}&=&(\delta_3 - \delta'_3)-\pi-3(\delta_1 - \delta_2)
\label{B8p}\\
\delta_{13}^{E_L}&=&(\delta_3 - \delta''_3)+ (\delta_3 - \delta_1) - \pi-2(\delta_1 - \delta_2)
\label{B9p}\\
\delta_{12}^{E_L}&=&(\delta'_2 - \delta_2)+(\delta_3 - \delta_2)
\label{B10p}
\eea

The leading charged lepton corrections to the
MNS angles and phases are given from Eqs.\ref{chlep23}-\ref{chlep12}:
\bea
s_{23}e^{-i\delta_{23}}
& \approx &
e^{-i\delta_{23}^{\nu_L}}\left[ s_{23}^{\nu_L}
-\theta_{23}^{E_L}c_{23}^{\nu_L}e^{-i(\delta_{23}^{E_L}-\delta_{23}^{\nu_L})} \right]
\label{chlep23p}
\\
\theta_{13}e^{-i\delta_{13}}
& \approx &
-\theta_{12}^{E_L}s_{23}^{\nu_L}e^{-i(\delta_{23}^{\nu_L}+\delta_{12}^{E_L})}
\label{chlep13p}
\\
s_{12}e^{-i\delta_{12}}
& \approx &
e^{-i\delta_{12}^{\nu_L}}\left[ s_{12}^{\nu_L}
-\theta_{12}^{E_L}c_{23}^{\nu_L}c_{12}^{\nu_L}e^{-i(\delta_{12}^{E_L}-\delta_{12}^{\nu_L})}
\right]
\label{chlep12p}
\eea
where we have kept the leading charged lepton correction in each term.
From Eqs.\ref{chlep23p}-\ref{chlep12p}, we see that the lepton phases are
approximately given by:
\bea
\delta_{23}&\approx & \delta_{23}^{\nu_L}
\label{delta23}
\\
\delta_{13}&\approx & \delta_{23}^{\nu_L}+\delta_{12}^{E_L}+\pi
\label{delta13}
\\
\delta_{12}&\approx & \delta_{12}^{\nu_L}
\eea
and hence $\delta$, the MNS CP phase relevant for neutrino oscillations,
is given by
\beq
\delta = \delta_{13}-\delta_{23}-\delta_{12}\approx
\delta_{12}^{E_L}-\delta_{12}^{\nu_L} +\pi
\approx \delta'_2 - \delta_2 +\pi.
\label{MNSphase}
\eeq
It is remarkable to observe that the phase appearing in the
leading charged lepton correction to the
solar angle in Eq.\ref{chlep12p} is just equal to
$\delta - \pi$, where $\delta$ is the MNS phase.
From Eqs.\ref{chlep23p}-\ref{chlep12p}
and the phases in Eqs.~\ref{B5p}-\ref{B7p} and Eqs.~\ref{B8p}-\ref{B10p}
and the tri-bimaximal neutrino mixing angles in Eqs.~\ref{23p}-\ref{13p},
the lepton mixing angles are given by:
\bea
s_{23}
& \approx &
%\frac{1}{\sqrt{2}}\left|1 +\theta_{23}^{E_L}e^{-i(\delta_3 - \delta'_3)} \right| \approx
\frac{1}{\sqrt{2}}\left( 1+ \theta_{23}^{E_L}\cos (\delta_3 - \delta'_3)  \right)
\label{chlep23pp}
\\
\theta_{13}
& \approx &
\frac{\theta_{12}^{E_L}}{\sqrt{2}}
\label{chlep13pp}
\\
s_{12}
& \approx &
%\frac{1}{\sqrt{3}}\left|1 - \theta_{12}^{E_L}e^{-i\delta} \right|
\frac{1}{\sqrt{3}}\left( 1- \theta_{12}^{E_L}\cos (\delta - \pi)  \right)
\label{chlep12pp}
\eea

We now turn to the quark sector.
The quark Yukawa matrices in Eqs.\ref{Yu},\ref{Yd} lead to
quark masses in the ratios:
\bea
m_d:m_s:m_b & \approx & \frac{|y_1||y_2|}{|y'_2|}\bar{\epsilon}^4:|y'_2|\bar{\epsilon}^2:y_{33}
\label{downmasses}
\\
m_u:m_c:m_t & \approx & \frac{|y_1||y_2|}{2|y'_2|}{\epsilon}^4:2|y'_2|{\epsilon}^2:y_{33}
\label{upmasses}
\eea
Comparing the down masses in Eq.\ref{downmasses} to
the charged lepton masses in Eq.\ref{chlepmasses} we see the expected
GJ relations (valid at the GUT scale):
\beq
\frac{m_e}{m_d}=\frac{1}{3}, \ \  \frac{m_\mu}{m_s}=3, \ \ \frac{m_\tau}{m_b}=1 .
\label{GJmasses}
\eeq
Using the conventions
in Appendix A, the quark Yukawa matrices in Eqs.\ref{Yu},\ref{Yd} lead to mixing angles:
\begin{eqnarray}
\theta^{U_L}_{23} & \approx & \frac{0.34|y'_3|{\epsilon}^2}{y_{33}}
\label{23U}\\
\theta^{U_L}_{13} & \approx & \frac{|y''_3|{\epsilon}^3}{y_{33}}
\label{13U} \\
\theta^{U_L}_{12} & \approx & \frac{|y_2|{\epsilon}}{2|y'_2|}
\label{12U}
\end{eqnarray}
\begin{eqnarray}
\theta^{D_L}_{23} & \approx & \frac{|y'_3|\bar{\epsilon}^2}{y_{33}}
\label{23D}\\
\theta^{D_L}_{13} & \approx & \frac{|y''_3|\bar{\epsilon}^3}{y_{33}}
\label{13D} \\
\theta^{D_L}_{12} & \approx & \frac{|y_2|\bar{\epsilon}}{|y'_2|}
\label{12D}
\end{eqnarray}
where we have written $y_{33}=|y_3|\bar{\epsilon}^{\frac{1}{2}}$,
and we have used a small quark angle approximation.
The auxiliary quark phases, used to make the quark mixing angles
real and positive, are exactly the same as the charged lepton phases
in Eq.\ref{tribiphasesE}, except that $\chi^{U_L}$ has an additional
phase of $\pi$ resulting from the negative 22 element of the up quark
Yukawa matrix. Using the results in Appendix A we find the CKM angles and phase:
\bea
\theta_{23}^{\mathrm{CKM}}
& \approx &
\theta_{23}^{D_L}-\theta_{23}^{U_L}
\label{quark23}
\\
\theta_{13}^{\mathrm{CKM}}
& \approx &
\theta_{13}^{D_L}
\label{quark13}
\\
\theta_{12}^{\mathrm{CKM}}
& \approx &
\theta_{12}^{D_L}+\theta_{12}^{U_L}
\label{quark12}
\\
\delta^{\mathrm{CKM}} & \approx & (\delta'_2 - \delta_2)+(\delta'_3 - \delta''_3)
\label{quarkphase}
\eea
The CKM phase is equal to the MNS phase in Eq.\ref{MNSphase} plus a second
independent phase determined by elements in the third column
of the Yukawa matrix which are irrelevant for neutrino mixing.

Since the CKM angles are approximately given by the down quark mixing angles,
using Eqs.\ref{quark23}-\ref{quark12}, Eqs.\ref{23D}-\ref{12D}, Eqs.\ref{23E}-\ref{12E},
we may relate the charged lepton mixing angles to the CKM angles,
\begin{eqnarray}
\theta^{E_L}_{23} & \approx & \theta^{D_L}_{23}
\approx \theta^{\mathrm{CKM}}_{23}
\label{23ED}\\
\theta^{E_L}_{13} & \approx & \theta^{D_L}_{13}
\approx \theta^{\mathrm{CKM}}_{13}
\label{13ED} \\
\theta^{E_L}_{12} & \approx & \frac{1}{3}\theta^{D_L}_{12}
\approx \frac{1}{3}\theta^{\mathrm{CKM}}_{12}
\approx \frac{1}{3}\theta_{C}
\label{12ED}
\end{eqnarray}
where the factor of $1/3$ in Eq.\ref{12ED} originates from the GJ structure.

Using Eqs.\ref{23ED}-\ref{12ED}, the lepton mixing angle relations in
Eqs.\ref{chlep23pp}-\ref{chlep12pp} become
\bea
\theta_{13}
& \approx &
\frac{\theta_{C}}{3\sqrt{2}}
\label{13predic}\\
s_{12}
& \approx &
%\frac{1}{\sqrt{3}}\left|1 - \theta_{12}^{E_L}e^{-i\delta} \right|
\frac{1}{\sqrt{3}}\left( 1- \frac{1}{3}\theta_{C}\cos (\delta - \pi)  \right)
\label{chlep12ppp}
\\
s_{23}
& \approx &
%\frac{1}{\sqrt{2}}\left|1 +\theta_{23}^{E_L}e^{-i(\delta_3 - \delta'_3)} \right| \approx
\frac{1}{\sqrt{2}}\left( 1+ \theta_{23}^{\mathrm{CKM}}\cos (\delta_3 - \delta'_3)  \right).
\label{chlep23ppp}
\eea
Eq.\ref{13predic} gives a prediction for the reactor angle:
\beq
\theta_{13}
\approx 3.06^\circ, \ \ \sin \theta_{13}\approx 0.052, \ \
\sin^2\theta_{13}\approx 2.7\times 10^{-3}, \ \
\sin^22\theta_{13}\approx 1.1\times 10^{-2}
\label{13predicnum}
\eeq
%\theta_{23} - {\theta_{23}^{\mathrm{CKM}}}
% \cos (\delta_3 - \delta'_3) & \approx & 45^\circ .
%\label{sum1p}
From Eqs.\ref{chlep12ppp},\ref{chlep23ppp}
the deviations from tri-bimaximal mixing may be expressed as:
\bea
\left| s_{12}^2- 1/3 \right| & \approx & \left|(2/9)\theta_{C}\cos \delta \right|<0.050
\label{dev12}
\\
\left| s_{23}^2-1/2 \right| & \approx &
\left| \theta_{23}^{\mathrm{CKM}}\cos (\delta_3 - \delta'_3) \right|<0.042
\label{dev23}
\eea
Eq.\ref{chlep23ppp} may also be expressed as
\beq
\theta_{23}^\circ \approx 45^\circ
+ {\theta_{23}^{\mathrm{CKM}}}^\circ \cos (\delta_3 - \delta'_3)
\label{dev23p}
\eeq
which shows that the atmospheric angle is maximal
$\theta_{23}= 45^\circ$ up to a correction
no larger than $\theta_{23}^{\mathrm{CKM}}\approx 2.4^\circ$
\beq
\theta_{23}= 45^\circ \pm 2.4^\circ .
\eeq
Eq.\ref{chlep12ppp} leads to
the tri-bimaximal complementarity relation in Eq.\ref{Eq:TBQLCrelation}:
\begin{eqnarray}
\theta_{12}^\circ + \frac{\theta_{C}^\circ }{3\sqrt{2}}\cos (\delta - \pi)  & \approx & 35.26^\circ
\label{sum2p}
\end{eqnarray}
Eq.\ref{sum2p} shows that the solar angle
takes its tri-bimaximal value $\theta_{12}=35.26^\circ$ up to a correction
no larger than $\frac{1}{\sqrt{2}}\frac{\theta_{\mathrm{C}}}{3}\approx 3.06^\circ$,
\beq
\theta_{12}=35.26^\circ \pm 3.06^\circ .
\label{solarerror}
\eeq
From Eq.\ref{13predic} and Eq.\ref{sum2p} we find the sum rule:
\begin{eqnarray}
\theta_{12}^\circ + \theta_{13}^\circ \cos (\delta - \pi)  & \approx & 35.26^\circ
\label{sum2pp}
\end{eqnarray}
Using Eq.\ref{chlep12ppp} we can
predict the neutrino oscillation phase $\delta$
from a future accurate measurement of the solar angle $\theta_{12}$:
\begin{eqnarray}
\cos (\delta - \pi) \approx 13.3(1-\sqrt{3}s_{12})
\label{MNSpredic}
\end{eqnarray}
or alternatively from Eq.\ref{sum2pp},
\begin{eqnarray}
\cos (\delta - \pi)  & \approx & \frac{35.26^\circ - \theta_{12}^\circ }{3.06^\circ}
\label{MNSpredic2}
\end{eqnarray}
For example, from an accurate measurement of the
solar angle of $\theta_{12}=33^\circ$ we predict
$\cos (\delta - \pi) = 0.74$ or $\delta =222^\circ$.

The above results are subject to some theoretical corrections as follows.
The renormalisation group running corrections
in running from the GUT scale $M_X$ to $M_Z$
depend strongly on $\tan \beta$ but may be typically estimated
for $\tan \beta \sim 40$ as
\footnote{These estimates have been made using the software
packages REAP/MPT introduced in \cite{Antusch:2005gp}}:
\bea
\theta_{12}(M_Z)-\theta_{12}(M_X) & \sim &  1^\circ \\
\theta_{13}(M_Z)-\theta_{13}(M_X) & \sim &  -0.5^\circ \\
\theta_{23}(M_Z)-\theta_{23}(M_X) & \sim &  2^\circ
\eea
In addition there is some theoretical error in the predictions of a similar
magnitude due to the analytic formulae used, the small angle approximations,
and the subleading operator corrections.

\section{Conclusions}
The poorly determined MNS parameters, when compared to the
accuracy of the measured quark mixing angles, presents an opportunity to
make testable predictions in the lepton sector by relating
the lepton mixing parameters to the quark ones.
In this paper we have shown how the neutrino mixing angles and oscillation phase
can be predicted from tri-bimaximal neutrino mixing,
corrected by charged lepton mixing angles which we relate
to quark mixing angles via quark-lepton unification.
The resulting predictions provide a probe of the high energy
structure of unified theories.

We have shown how tri-bimaximal neutrino mixing can originate
from the see-saw mechanism using sequential dominance.
We gave the conditions for
tri-bimaximal neutrino mixing to originate from
sequential dominance, thereby providing a general
and natural framework for this approach called
constrained sequential dominance (CSD).
We discussed a realisation of CSD based on
$SO(3)$ family symmetry and vacuum alignment,
although there are other examples that are possible.
We then constructed a realistic model of
quark and lepton masses and mixings based on the $SO(3)$ family
symmetry and vacuum alignment, together with
quark-lepton unification arising from a Pati-Salam gauge group.
With the ingredients of tri-bimaximal complementarity in place,
the MNS parameters were then predicted in terms of the CKM
parameters. Although these predictions have been derived for the
specific model presented, we would expect them to apply to a more
general class of models based on real
vacuum alignment which lead to tri-bimaximal complementarity.

The atmospheric angle is predicted to be approximately maximal
$\theta_{23}= 45^\circ$, corrected by
the quark mixing angle $\theta_{23}^{\mathrm{CKM}}\approx 2.4^\circ$,
with the correction controlled by an undetermined phase in the quark sector.
The solar angle is predicted by the tri-bimaximal complementarity
relation: $\theta_{12}+ \frac{1}{\sqrt{2}}\frac{\theta_{\mathrm{C}}}{3}
\cos (\delta - \pi) \approx 35.26^\circ $, where $\theta_{\mathrm{C}}$ is the
Cabibbo angle and $\delta$ is the neutrino oscillation phase.
The reactor angle is predicted to be
$\theta_{13} \approx \frac{1}{\sqrt{2}}\frac{\theta_{\mathrm{C}}}{3}\approx 3.06^\circ$.
The neutrino oscillation phase $\delta$ is predicted in terms of the solar angle
to be $\cos (\delta - \pi) \approx  (35.26^\circ - \theta_{12}^\circ )/3.06^\circ $.
These predictions can all be tested by future high precision
neutrino oscillation experiments.
Indeed the link between low energy neutrino parameters and quark-lepton unification
provides a powerful theoretical motivation for performing
high precision neutrino oscillation experiments.
In particular the prediction of the neutrino oscillation phase in terms
of the solar angle is a remarkable result, which motivates
an accurate measurement of the solar angle.
The theoretical prediction for the reactor angle could be tested
with the next generation of superbeam or reactor experiments, and
the prediction for the oscillation phase
could be accurately tested at a Neutrino Factory.

\section*{Acknowledgements}
I would like to thank Stefan Antusch,
for helpful discussions.

\section*{Appendix}
\appendix

\renewcommand{\thesection}{\Alph{section}}
\renewcommand{\thesubsection}{\Alph{section}.\arabic{subsection}}
\def\theequation{\Alph{section}.\arabic{equation}}
\renewcommand{\thetable}{\arabic{table}}
\renewcommand{\thefigure}{\arabic{figure}}
\setcounter{section}{0}
\setcounter{equation}{0}

\section{Conventions and Mixing Formalism}\label{conventions}
We shall use the conventions defined in \cite{King:2002nf}.
The Dirac mass matrices of the charged leptons
and neutrinos are given by $m^{\mathrm{E}}_{LR}=Y^\mathrm{E}_{LR}v_\mathrm{d}$,
and $m^{\nu}_{LR}=Y^{\nu}_{LR}v_\mathrm{u}$
where
$v_\mathrm{d} = \< h^0_\mathrm{d}\>$ and $v_\mathrm{u} = \<
h^0_\mathrm{u}\>$,
and the Lagrangian is of the form ${\cal L}=-\bar{\psi}_LY_{LR}h\psi_R +H.c.$
The neutrino mass
matrix $m^\nu_{LL}$ is given by the type I see-saw mechanism as
\beq
m_{LL}^{\nu}=m_{LR}^{\nu}M_{RR}^{-1}{m_{LR}^{\nu \ T}},
\label{seesaw}
\eeq
in terms of the Dirac neutrino mass matrix $m_{LR}^{\nu}$ and the heavy Majorana
mass matrix $M_{RR}$. In this convention the effective Majorana
masses are given by the Lagrangian
${\cal L}=-\bar{\nu}_Lm^{\nu}_{LL}\nu^c +H.c.$
The change from flavour basis to mass
eigenbasis can be performed with the unitary diagonalization matrices
$V_{E_\mathrm{L}},V_{E_\mathrm{R}}$ and
$V_{\nu_\mathrm{L}}$ by
\begin{eqnarray}\label{eq:DiagMe}
V_{E_\mathrm{L}} \, m^{\mathrm{E}}_{LR} \,V^\dagger_{E_\mathrm{R}} =
\left(\begin{array}{ccc}
\!m_e&0&0\!\\
\!0&m_\mu&0\!\\
\!0&0&m_\tau\!
\end{array}
\right)\! , \quad
V_{\nu_\mathrm{L}} \,m^\nu_{\mathrm{LL}}\,V^T_{\nu_\mathrm{L}} =
\left(\begin{array}{ccc}
\!m_1&0&0\!\\
\!0&m_2&0\!\\
\!0&0&m_3\!
\end{array}
\right)\! .
\end{eqnarray}
The MNS matrix is then given by
\begin{eqnarray}\label{Eq:MNS_Definition}
U_{\mathrm{MNS}} = V_{e_\mathrm{L}} V^\dagger_{\nu_\mathrm{L}}\; .
\end{eqnarray}
We use the parameterization
$
U_{\mathrm{MNS}} = U_{23} U_{13} U_{12}
$
with $U_{23}, U_{13}, U_{12}$ being defined as
\begin{align}\label{eq:U23U13U12}
U_{12}=
\left(\begin{array}{ccc}
  c_{12} & s_{12}e^{-i\delta_{12}} & 0\\
  -s_{12}e^{i\delta_{12}}&c_{12} & 0\\
  0&0&1\end{array}\right)
  , \:&
\quad U_{13}=\left(\begin{array}{ccc}
   c_{13} & 0 & s_{13}e^{-i\delta_{13}}\\
  0&1& 0\\
  - s_{13}e^{i\delta_{13}}&0&c_{13}\end{array}\right)  ,  \nonumber \\
U_{23}=\left(\begin{array}{ccc}
 1 & 0 & 0\\
0&c_{23} & s_{23}e^{-i\delta_{23}}\\
0&-s_{23}e^{i\delta_{23}}&c_{23}
 \end{array}\right)
\end{align}
where $s_{ij}$ and $c_{ij}$ stand for $\sin (\theta_{ij})$ and $\cos
(\theta_{ij})$, respectively.
$\delta$, the Dirac CP phase relevant for neutrino oscillations,
is given by $\delta = \delta_{13}-\delta_{23}-\delta_{12}$.

The MNS matrix is thus constructed as a product
of a unitary matrix from the charged lepton sector $V^{E_L}$
and a unitary matrix from the neutrino sector ${V^{\nu_L}}^{\dagger}$.
Each of these unitary matrices may be parametrised as:
\beq
V^{\dagger}=P_2R_{23}R_{13}P_1R_{12}P_3
\label{V1}
\eeq
where $R_{ij}$ are a sequence of real rotations corresponding to the
Euler angles $\theta_{ij}$, and $P_i$ are diagonal phase matrices.
The Euler matrices are given by
\begin{equation}
R_{23}=
\left(\begin{array}{ccc}
1 & 0 & 0 \\
0 & c_{23} & s_{23} \\
0 & -s_{23} & c_{23} \\
\end{array}\right)
\label{R23}
\end{equation}
\begin{equation}
R_{13}=
\left(\begin{array}{ccc}
c_{13} & 0 & s_{13} \\
0 & 1 & 0 \\
-s_{13} & 0 & c_{13} \\
\end{array}\right)
\label{R13}
\end{equation}
\begin{equation}
R_{12}=
\left(\begin{array}{ccc}
c_{12} & s_{12} & 0 \\
-s_{12} & c_{12} & 0\\
0 & 0 & 1 \\
\end{array}\right)
\label{R12}
\end{equation}
where $c_{ij} = \cos\theta_{ij}$ and $s_{ij} = \sin\theta_{ij}$.
The phase matrices are given by
\beq
P_1=
\left( \begin{array}{ccc}
1 & 0 & 0    \\
0 & e^{i\chi} & 0 \\
0 & 0 & 1
\end{array}
\right)
\label{P1}
\eeq
\beq
P_2=
\left( \begin{array}{ccc}
1 & 0 & 0    \\
0 & e^{i\phi_2} & 0 \\
0 & 0 & e^{i\phi_3}
\end{array}
\right)
\label{P2}
\eeq
\beq
P_3=
\left( \begin{array}{ccc}
e^{i\omega_1} & 0 & 0    \\
0 & e^{i\omega_2} & 0 \\
0 & 0 & e^{i\omega_3}
\end{array}
\right)
\label{P3}
\eeq
Thus we write
\beq
{V^{\nu_L}}^{\dagger}
=P_2^{\nu_L}R_{23}^{\nu_L}R_{13}^{\nu_L}P_1^{\nu_L}R_{12}^{\nu_L}P_3^{\nu_L}
\label{VnuL}
\eeq
\beq
{V^{E_L}}^{\dagger}
=P_2^{E_L}R_{23}^{E_L}R_{13}^{E_L}P_1^{E_L}R_{12}^{E_L}P_3^{E_L}
\label{VEL}
\eeq
in terms of independent angles
and phases for the left-handed neutrino and charged lepton sectors
distinguished by the superscripts $\nu_L$ and $E_L$.

The MNS matrix can be expanded
in terms of neutrino and charged lepton mixing angles and phases
to leading order in the charged lepton mixing angles which are assumed
small:
\footnote{Note that the sign of the last term in Eq.\ref{chlep13} is reversed
compared to the results quoted in \cite{King:2002nf}. I am grateful
to Stefan Antusch for correcting these results.}
\bea
s_{23}e^{-i\delta_{23}}
& \approx &
s_{23}^{\nu_L}e^{-i\delta_{23}^{\nu_L}}
-\theta_{23}^{E_L}
c_{23}^{\nu_L}e^{-i\delta_{23}^{E_L}}
\label{chlep23}
\\
\theta_{13}e^{-i\delta_{13}}
& \approx &
\theta_{13}^{\nu_L}e^{-i\delta_{13}^{\nu_L}}
-\theta_{13}^{E_L}c_{23}^{\nu_L}e^{-i\delta_{13}^{E_L}}
-\theta_{12}^{E_L}s_{23}^{\nu_L}e^{i(-\delta_{23}^{\nu_L}-\delta_{12}^{E_L})}
\label{chlep13}
\\
s_{12}e^{-i\delta_{12}}
& \approx &
s_{12}^{\nu_L}e^{-i\delta_{12}^{\nu_L}}
+\theta_{13}^{E_L}
c_{12}^{\nu_L}s_{23}^{\nu_L}e^{i(\delta_{23}^{\nu_L}-\delta_{13}^{E_L})}
- \theta_{12}^{E_L} c_{23}^{\nu_L}c_{12}^{\nu_L}e^{-i\delta_{12}^{E_L}}
\label{chlep12}
\eea
where
\bea
\delta_{12}^{\nu_L}&=&\omega_1^{\nu_L}-\omega_2^{\nu_L}
\label{B5}\\
\delta_{13}^{\nu_L}&=&\omega_1^{\nu_L}-\omega_3^{\nu_L}
\label{B6}\\
\delta_{23}^{\nu_L}&=&\chi^{\nu_L}+\omega_2^{\nu_L}-\omega_3^{\nu_L}
\label{B7}\\
\delta_{23}^{E_L}&=&
-\phi_2^{E_L}+\phi_3^{E_L}+\phi_2^{\nu_L}-\phi_3^{\nu_L}
+\chi^{\nu_L}+\omega_2^{\nu_L}-\omega_3^{\nu_L}
\label{B8}\\
\delta_{13}^{E_L}&=&\phi_3^{E_L}-\phi_3^{\nu_L}
+\omega_1^{\nu_L}-\omega_3^{\nu_L}
\label{B9}\\
\delta_{12}^{E_L}&=&\chi^{E_L}+\phi_2^{E_L}-\phi_2^{\nu_L}
-\chi^{\nu_L}+\omega_1^{\nu_L}-\omega_2^{\nu_L}
\label{B10}
\eea

In the quark sector an analagous procedure is followed.
The Dirac mass matrices of the quarks
are given by $m^{\mathrm{D}}_{LR}=Y^\mathrm{D}_{LR}v_\mathrm{d}$,
and $m^{\mathrm{U}}_{LR}=Y^{\mathrm{U}}_{LR}v_\mathrm{u}$.
The change from flavour basis to mass
eigenbasis can be performed with the unitary diagonalization matrices
$V_{D_\mathrm{L}},V_{D_\mathrm{R}}$ and
$V_{U_\mathrm{L}},V_{U_\mathrm{R}}$
by
\begin{eqnarray}\label{eq:DiagMq}
V_{D_\mathrm{L}} \, m^{\mathrm{D}}_{LR} \,V^\dagger_{D_\mathrm{R}} =
\left(\begin{array}{ccc}
\!m_d&0&0\!\\
\!0&m_s&0\!\\
\!0&0&m_b\!
\end{array}
\right)\! ,
\quad
V_{U_\mathrm{L}} \, m^{\mathrm{U}}_{LR} \,V^\dagger_{U_\mathrm{R}} =
\left(\begin{array}{ccc}
\!m_u&0&0\!\\
\!0&m_c&0\!\\
\!0&0&m_t\!
\end{array}
\right)\! ,
\end{eqnarray}
The CKM matrix is then given by
\begin{eqnarray}\label{Eq:CKM_Definition}
U_{\mathrm{CKM}} = V_{U_\mathrm{L}} V^\dagger_{D_\mathrm{L}}.
\end{eqnarray}
We use the standard parameterization
$
U_{\mathrm{CKM}} = R^{\mathrm{CKM}}_{23} U^{\mathrm{CKM}}_{13} R^{\mathrm{CKM}}_{12}
$
where we label the quark parameters as CKM to distinguish them from the
(unlabelled) lepton mixing angles. If the CKM angles are given predominantly by the down
mixing angles, then we may use the analagous results to those quoted above to obtain the
corrections coming from the up sector. Thus for example the analagous relations to
Eq.\ref{chlep23}-\ref{B10} apply in the quark sector also, with the replacements
$\nu \rightarrow D$ and $E \rightarrow U$.
In the quark sector the phases
$\omega^{D_L}_i$, $\omega^{U_L}_i$
are all undetermined and are used to remove phases from the MNS matrix.
In particular $\omega^{D_L}_i$ may be used to set the
phases $\delta_{12}^{CKM}=\delta_{23}^{CKM}=0$,
with a single CKM phase remaining,
$\delta^{\mathrm{CKM}}=\delta_{13}^{\mathrm{CKM}}$.

\section{Vacuum Alignment}\label{vacuum}
In order to achieve the desired vacuum alignment in this model , we shall
introduce the following additional superpotential terms:
\begin{eqnarray}
W_{{\rm SB}} &\sim &
A(\phi_{1}^2-\Lambda_1^2)+B(\phi_{2}^2-\Lambda_2^2)+C(\phi_{3}^2-\Lambda_3^2) \label{ABC}
\\
&+&
D\phi_{1}.\phi_{2}+E\phi_{1}.\phi_{3}+F\phi_{2}.\phi_{3} \label{DEF}
\\
%&+&
%G(\phi_{12}^2-\Lambda_4^2)+H(\tilde{\phi}_{12}^2-\Lambda_5^2)\label{GH}
%\\
%&+&
%I(\phi_{23}^2-\Lambda_6^2)+J(\tilde{\phi}_{23}^2-\Lambda_7^2)\label{IJ}
%\\
%&+&
%K(\phi_{123}^2-\Lambda_8^2)\label{K}
%\\
&+&
L\phi_{12}.\tilde{\phi}_{12}+M\phi_{23}.\tilde{\phi}_{23}\label{LM}
\\
&+&
N\phi_{123}.{\phi}_{12}+O\phi_{123}.{\phi}_{23}\label{NO}
\\
&+&
P((\phi_{12}.\phi_{1})(\phi_{12}.\phi_{2})-\Lambda_9^2)
+Q((\tilde{\phi}_{12}.\phi_{1})(\tilde{\phi}_{12}.\phi_{2})-\Lambda_{10}^2)
\label{PQ}
\\
&+&
R((\phi_{23}.\phi_{2})(\phi_{23}.\phi_{3})-\Lambda_{11}^2)
+S((\tilde{\phi}_{23}.\phi_{2})(\tilde{\phi}_{23}.\phi_{3})-\Lambda_{12}^2)
\label{RS}
\\
&+&
T((\phi_{123}.\phi_{1})(\phi_{123}.\phi_{2})(\phi_{123}.\phi_{3})-\Lambda_{13}^2)
\label{T}
\end{eqnarray}
where $A,\cdots T$ are $SO(3)$ and Pati-Salam singlet superfields and
$\Lambda_i$ are independent heavy mass scales which we regard as
arising from the VEVs of some $SO(3)$ singlet fields.
Such VEVs could arise from some radiative symmetry breaking
mechanism, for example \cite{GG}.
$\Lambda_i^2$ can be taken to be real and positive by a suitable phase
choice for the fields $A,\cdots T$ \cite{Barbieri:1999km}.
Note that in such an $SO(3)$ theory
with real VEVs all the D-terms will be automatically zero,
and the vacuum alignment is then achieved purely from the F-terms
being minimised to zero, up to soft supersymmetry breaking perturbations.
It is straightforward to deduce the required quantum numbers of these
superfields under the symmetry group $R\times Z_4^2\times Z_3^2\times
U(1)$ from the quantum number assignments of the flavons in Table 1,
and the requirement that the superpotential terms given above are
allowed.

The potential
consists of F-terms of the form $|F_X|^2$, together with positive
soft mass squareds for the flavon fields.
Since $\Lambda_i^2$ are real and positive this results in real
VEVs as discussed in \cite{Antusch:2004xd,Barbieri:1999km},
which greatly simplifies the analysis, and crucially
restricts the number of undetermined phases in the analysis, ultimately
leading to a prediction for the neutrino oscillation phase.
The purpose of the terms in Eq.\ref{ABC},\ref{DEF} is for the F-terms
$|F_X|^2$, with $X=A,\cdots F$, to be minimised by real three orthogonal VEVs
for $\phi_{1,2,3}$ of the form given in Eq.\ref{123}.
In particular the terms in Eq.\ref{ABC} drive the VEVs
to be non-zero, and the the terms in Eq.\ref{DEF} together with the
soft positive mass squareds
then lead to real and orthogonal VEVs
of the form given in Eq.\ref{123}.
The purpose of the remaining terms is to align the VEVs of the remaining
fields relative to these basis vectors, in order to achieve the alignment
for $\phi_{23},\phi_{123}$ as shown in Eq.\ref{123}, as follows.

To achieve the alignment of $\phi_{23}$ requires two fields
$\phi_{23},\tilde{\phi}_{23}$. The purpose of the terms in
Eq.\ref{RS} is to drive non-zero VEVs for these two fields
in the $(2,3)$ directions, and taken together with the positive
soft mass squared terms,
\footnote{I am grateful to Graham Ross and Ivo de Medeiros-Varzielas
for suggesting the use of soft mass terms to achieve this alignment.
The use of soft mass terms for alignment is also discussed in \cite{GG}.}
lead to a potential
which is minimised when magnitudes of the VEVs along each of the
directions is equal, for each of the two fields
$\phi_{23},\tilde{\phi}_{23}$ separately.
This is because for each of these flavons the potential takes the form
$V=m_{soft}^2(y^2+z^2)+b^2(yz-M_2^2)^2$,
where all parameters are real and positive and $y,z$ are the
components of the VEVs along the $2,3$ directions respectively.
Such a potential is minimised by component VEVs of equal magnitude $y^2=z^2$.
The term in Eq.\ref{LM} proportional to $M$
then ensures that the two VEVs are orthogonal to each other,
and without loss of generality this results in VEVs of the form:
\beq \label{23pp}
\begin{array}{l}
\tilde{\phi}_{23}\sim
\begin{pmatrix}
0 \\
1 \\
1 \\
\end{pmatrix}, \ \ \;
\phi_{23}\sim
\begin{pmatrix}
0 \\
1 \\
-1 \\
\end{pmatrix}.
\end{array}
\eeq
Following analagous arguments, the terms in
Eq.\ref{PQ}, together with positive soft mass terms and
the term proportional to $L$
in Eq.\ref{LM} leads, without loss of generality, to VEVs
of the form:
\beq \label{12pp}
\begin{array}{l}
\tilde{\phi}_{12}\sim
\begin{pmatrix}
1 \\
1 \\
0 \\
\end{pmatrix}, \ \ \;
\phi_{12}\sim
\begin{pmatrix}
1 \\
-1 \\
0 \\
\end{pmatrix}.
\end{array}
\eeq

The alignment of $\phi_{123}$ is achieved by the term in Eq.\ref{T}
which drives the VEV, and ensures that
all three components of the VEV are non-zero,
and taken together the
soft mass terms and F-terms in the potential which
result from these terms imply that the
component VEVs must have equal magnitude.
The terms in Eq.\ref{NO} then align these
components to be orthogonal to both $\phi_{12}$
and $\phi_{23}$, resulting in the alignment assumed in
Eq.\ref{123}:
\beq \label{123p}
\begin{array}{l}
\phi_{123}\sim
\begin{pmatrix}
1 \\
1 \\
1 \\
\end{pmatrix}.
\end{array}
\eeq

\providecommand{\bysame}{\leavevmode\hbox to3em{\hrulefill}\thinspace}


\begin{thebibliography}{1}

%\cite{Maltoni:2004ei}
\bibitem{Maltoni:2004ei}
M.~Maltoni, T.~Schwetz, M.~A.~Tortola and J.~W.~F.~Valle,
%``Status of global fits to neutrino oscillations,''
New J.\ Phys.\  {\bf 6} (2004) 122
[arXiv:hep-ph/0405172].
%%CITATION = HEP-PH 0405172;%%

\bibitem{RPP}
F.J.Gilman et al, in {\it Review of Particle Physics} (2004), p.130.

%\cite{King:2003jb}
\bibitem{King:2003jb}
For a review see e.g. S.~F.~King,
%``Neutrino mass models,''
Rept.\ Prog.\ Phys.\  {\bf 67} (2004) 107 [arXiv:hep-ph/0310204].
%%CITATION = HEP-PH 0310204;%%

%\cite{Minakata:2005rf}
\bibitem{Minakata:2005rf}
H.~Minakata,
``Quark-Lepton Complementarity: a Review,''
arXiv:hep-ph/0505262.
%%CITATION = HEP-PH 0505262;%%

\bibitem{Raidal:2004iw}
  M.~Raidal,
  %``Prediction Theta(c) + Theta(sol) = pi/4 from flavor physics: A new evidence
  %for grand unification?,''
  Phys.\ Rev.\ Lett.\  {\bf 93} (2004) 161801, hep-ph/0404046.
  %%CITATION = HEP-PH 0404046;%%

\bibitem{Minakata:2004xt}
  H.~Minakata and A.~Y.~Smirnov,
  %``Neutrino mixing and quark lepton complementarity,''
  Phys.\ Rev.\ D {\bf 70}, 073009 (2004), hep-ph/0405088.
  %%CITATION = HEP-PH 0405088;%%

\bibitem{mohfram} %\bibitem{Frampton:2004vw}
  P.~H.~Frampton and R.~N.~Mohapatra,
  %``Possible gauge theoretic origin for quark-lepton complementarity,''
  JHEP {\bf 0501}, 025 (2005), hep-ph/0407139.
  %%CITATION = HEP-PH 0407139;%%


\bibitem{QLCliterature}
  J.~Ferrandis and S.~Pakvasa,
  %``Quark-lepton complementarity relation and neutrino mass hierarchy,''
  Phys.\ Rev.\ D {\bf 71} (2005) 033004,
  hep-ph/0412038;
  %%CITATION = HEP-PH 0412038;%%
%\bibitem{Kang:2005as}
 S.~K.~Kang, C.~S.~Kim and J.~Lee,
  %``Quark-lepton complementarity with renormalization effects through threshold
  %corrections,''
  arXiv:hep-ph/0501029;
  %%CITATION = HEP-PH 0501029;%%
%\bibitem{Li:2005ir}
  N.~Li and B.~Q.~Ma,
  %``Unified parametrization of quark and lepton mixing matrices,''
  hep-ph/0501226;
  %%CITATION = HEP-PH 0501226;%%
%%%
  K.~Cheung, S.~K.~Kang, C.~S.~Kim and J.~Lee,
  %``Lepton flavor violation as a probe of quark-lepton unification,''
  hep-ph/0503122;
  %%CITATION = HEP-PH 0503122;%%
  %\bibitem{Xing:2005ur}
  Z.~z.~Xing,
  %``Nontrivial correlation between the CKM and MNS matrices,''
  %arXiv:
  hep-ph/0503200;
  %%CITATION = HEP-PH 0503200;%%
%\bibitem{Datta:2005ci}
  A.~Datta, L.~Everett and P.~Ramond,
  %``Cabibbo haze in lepton mixing,''
  arXiv:
  hep-ph/0503222;
  %%CITATION = HEP-PH 0503222;%%
%\cite{Ohlsson:2005js}
%\bibitem{Ohlsson:2005js}
T.~Ohlsson,
%``Bimaximal fermion mixing from the quark and leptonic mixing matrices,''
arXiv:hep-ph/0506094.
%%CITATION = HEP-PH 0506094;%%



%\cite{Antusch:2005ca}
\bibitem{Antusch:2005ca} S.~Antusch, S.~F.~King and R.~N.~Mohapatra,
%``Quark-Lepton Complementarity in Unified Theories,''
arXiv:hep-ph/0504007.
%%CITATION = HEP-PH 0504007;%%

%\cite{Lindner:2005pk}
\bibitem{Lindner:2005pk}
M.~Lindner, M.~A.~Schmidt and A.~Y.~Smirnov,
%``Screening of Dirac flavor structure in the seesaw and neutrino mixing,''
arXiv:hep-ph/0505067.
%%CITATION = HEP-PH 0505067;%%


%\cite{King:2000ce}
\bibitem{King:2000ce}
S.~F.~King and N.~N.~Singh,
%``Inverted hierarchy models of neutrino masses,''
Nucl.\ Phys.\ B {\bf 596} (2001) 81
[arXiv:hep-ph/0007243].
%%CITATION = HEP-PH 0007243;%%

%\cite{King:2002nf}
\bibitem{King:2002nf}
S.~F.~King,
%``Constructing the large mixing angle MNS matrix in see-saw models with
%right-handed neutrino dominance,''
JHEP {\bf 0209} (2002) 011
[arXiv:hep-ph/0204360];
%%CITATION = HEP-PH 0204360;%%
%\cite{King:2002qh}
%\bibitem{King:2002qh}
S.~F.~King,
%``Leptogenesis - MNS link in unified models with natural neutrino mass
%hierarchy,''
Phys.\ Rev.\ D {\bf 67} (2003) 113010
[arXiv:hep-ph/0211228].
%%CITATION = HEP-PH 0211228;%%


%\cite{Georgi:1979df}
\bibitem{Georgi:1979df}
H.~Georgi and C.~Jarlskog,
%``A New Lepton - Quark Mass Relation In A Unified Theory,''
Phys.\ Lett.\ B {\bf 86} (1979) 297.
%%CITATION = PHLTA,B86,297;%%

\bibitem{tribi}
%\cite{Harrison:2002er}
%bibitem{Harrison:2002er}
P.~F.~Harrison, D.~H.~Perkins and W.~G.~Scott,
%``Tri-bimaximal mixing and the neutrino oscillation data,''
Phys.\ Lett.\ B {\bf 530} (2002) 167
[arXiv:hep-ph/0202074];
%%CITATION = HEP-PH 0202074;%%
%\cite{Harrison:2002kp}
%\bibitem{Harrison:2002kp}
P.~F.~Harrison and W.~G.~Scott,
%``Symmetries and generalisations of tri-bimaximal neutrino mixing,''
Phys.\ Lett.\ B {\bf 535} (2002) 163
[arXiv:hep-ph/0203209];
%%CITATION = HEP-PH 0203209;%%
%\cite{Harrison:2003aw}
%\bibitem{Harrison:2003aw}
P.~F.~Harrison and W.~G.~Scott,
%``Permutation symmetry, tri-bimaximal neutrino mixing and the S3 group
%characters,''
Phys.\ Lett.\ B {\bf 557} (2003) 76
[arXiv:hep-ph/0302025];
%%CITATION = HEP-PH 0302025;%%
C.I.Low and R.R.Volkas, hep-ph/0305243;
see also: %\cite{Wolfenstein:1978uw}
%bibitem{Wolfenstein:1978uw}
L.~Wolfenstein,
%``Oscillations Among Three Neutrino Types And CP Violation,''
Phys.\ Rev.\ D {\bf 18} (1978) 958.
%%CITATION = PHRVA,D18,958;%%

%\cite{Ferrandis:2004mq}
\bibitem{Ferrandis:2004mq}
J. D. Bjorken(unpublished);
J.~Ferrandis and S.~Pakvasa,
%``A prediction for $|$U(e3)$|$ from patterns in the charged lepton spectra,''
Phys.\ Lett.\ B {\bf 603} (2004) 184
[arXiv:hep-ph/0409204].
%%CITATION = HEP-PH 0409204;%%


\bibitem{GG}
G.~G.~Ross, Proceedings of Seesaw 25, Int. Conf. on the Seesaw
mechanism, 10-11 June 2004, Paris;
%\cite{King:2003rf}
%\bibitem{King:2003rf}
S.~F.~King and G.~G.~Ross,
%``Fermion masses and mixing angles from SU(3) family symmetry and
%unification,''
Phys.\ Lett.\ B {\bf 574} (2003) 239
[arXiv:hep-ph/0307190];
%%CITATION = HEP-PH 0307190;%%
%\cite{King:2001uz}
%\bibitem{King:2001uz}
S.~F.~King and G.~G.~Ross,
%``Fermion masses and mixing angles from SU(3) family symmetry,''
Phys.\ Lett.\ B {\bf 520} (2001) 243
[arXiv:hep-ph/0108112].
%%CITATION = HEP-PH 0108112;%%

%\cite{Altarelli:2005yp}
\bibitem{Altarelli:2005yp}
G.~Altarelli and F.~Feruglio,
%``Tri-bimaximal neutrino mixing from discrete symmetry in extra dimensions,''
arXiv:hep-ph/0504165;
%%CITATION = HEP-PH 0504165;%%
%\cite{Ma:2005sh}
%\bibitem{Ma:2005sh}
E.~Ma,
%``Aspects of the tetrahedral neutrino mass matrix,''
arXiv:hep-ph/0505209;
%%CITATION = HEP-PH 0505209;%%
%\cite{Ma:2004yb}
%\bibitem{Ma:2004yb}
E.~Ma,
%``A4 symmetry and neutrinos with very different masses,''
Phys.\ Rev.\ D {\bf 70} (2004) 031901.
%%CITATION = PHRVA,D70,031901;%%



%\cite{King:1998jw}
\bibitem{King:1998jw}
S.~F.~King,
%``Atmospheric and solar neutrinos with a heavy singlet,''
Phys.\ Lett.\ B {\bf 439} (1998) 350
[arXiv:hep-ph/9806440];
%%CITATION = HEP-PH 9806440;%%
S.~F.~King,
%``Atmospheric and solar neutrinos from single right-handed neutrino  dominance
%and U(1) family symmetry,''
Nucl.\ Phys.\ B {\bf 562} (1999) 57
[arXiv:hep-ph/9904210];
%%CITATION = HEP-PH 9904210;%%
S.~F.~King,
%``Large mixing angle MSW and atmospheric neutrinos from single  right-handed
%neutrino dominance and U(1) family symmetry,''
Nucl.\ Phys.\ B {\bf 576} (2000) 85
[arXiv:hep-ph/9912492].
%%CITATION = HEP-PH 9912492;%%
%S.~F.~King,
%``Constructing the large mixing angle MNS matrix in see-saw models with
%right-handed neutrino dominance,''
%JHEP {\bf 0209} (2002) 011
%[arXiv:hep-ph/0204360].
%%CITATION = HEP-PH 0204360;%%

%\cite{Antusch:2004xd}
\bibitem{Antusch:2004xd}
S.~Antusch and S.~F.~King,
%``From hierarchical to partially degenerate neutrinos via type II upgrade of
%type I see-saw models,''
Nucl.\ Phys.\ B {\bf 705} (2005) 239
[arXiv:hep-ph/0402121].
%%CITATION = HEP-PH 0402121;%%

%\cite{Antusch:2004xy}
\bibitem{Antusch:2004xy}
%\cite{Hambye:2003ka}
%\bibitem{Hambye:2003ka}
T.~Hambye and G.~Senjanovic,
%``Consequences of triplet seesaw for leptogenesis,''
Phys.\ Lett.\ B {\bf 582} (2004) 73
[arXiv:hep-ph/0307237];
%%CITATION = HEP-PH 0307237;%%
S.~Antusch and S.~F.~King,
%``Type II leptogenesis and the neutrino mass scale,''
Phys.\ Lett.\ B {\bf 597} (2004) 199
[arXiv:hep-ph/0405093].
%%CITATION = HEP-PH 0405093;%%

%\cite{Barbieri:1999km}
\bibitem{Barbieri:1999km}
R.~Barbieri, L.~J.~Hall, G.~L.~Kane and G.~G.~Ross,
%``Nearly degenerate neutrinos and broken flavour symmetry,''
arXiv:hep-ph/9901228.
%%CITATION = HEP-PH 9901228;%%

\bibitem{Roberts:2001zy}
R.~G. Roberts, A.~Romanino, G.~G. Ross, and L.~Velasco-Sevilla,
%\emph{Precision
%  test of a fermion mass texture},
Nucl. Phys. \textbf{B615} (2001), 358--384,
  {hep-ph/0104088};
%\cite{Allanach:1995sj}
%\bibitem{Allanach:1995sj}
B.~C.~Allanach and S.~F.~King,
%``Fermion masses in a supersymmetric SU(4) x SU(2)-L x SU(2)-R model,''
Nucl.\ Phys.\ B {\bf 456} (1995) 57
[arXiv:hep-ph/9502219];
%%CITATION = HEP-PH 9502219;%%
%\cite{Allanach:1995ch}
%\bibitem{Allanach:1995ch}
B.~C.~Allanach and S.~F.~King,
%``Neutrino Masses and Mixing Angles in a Supersymmetric SU(4)$ \otimes
%$SU(2)$_L \otimes $SU(2)$_R$ Model,''
Nucl.\ Phys.\ B {\bf 459} (1996) 75
[arXiv:hep-ph/9509205];
%%CITATION = HEP-PH 9509205;%%
%\cite{Allanach:1997gu}
%\bibitem{Allanach:1997gu}
B.~C.~Allanach, S.~F.~King, G.~K.~Leontaris and S.~Lola,
%``Yukawa textures from family symmetry and unification,''
Phys.\ Lett.\ B {\bf 407} (1997) 275
[arXiv:hep-ph/9703361];
%%CITATION = HEP-PH 9703361;%%
%\bibitem{Allanach:1996hz}
B.~C. Allanach, S.~F. King, G.~K. Leontaris, and S.~Lola,
%\emph{A new approach
%  to {Y}ukawa textures in supersymmetric unified models with gauged family
%  symmetries},
Phys. Rev. \textbf{D56} (1997), 2632--2655,
  {hep-ph/9610517}.

\bibitem{ps} J. C. Pati and A. Salam, Phys. Rev. {\bf D 10}, 270 (1974).

%\cite{King:1994fv}
\bibitem{King:1994fv}
S.~F.~King,
%``SU(4) x SU(2)-L x SU(2)-R as a surrogate SUSY GUT,''
Phys.\ Lett.\ B {\bf 325} (1994) 129
[Erratum-ibid.\ B {\bf 325} (1994) 538];
%%CITATION = PHLTA,B325,129;%%

%\cite{King:1997ia}
\bibitem{King:1997ia}
S.~F.~King and Q.~Shafi,
%``Minimal supersymmetric SU(4) x SU(2)L x SU(2)R,''
Phys.\ Lett.\ B {\bf 422} (1998) 135
[arXiv:hep-ph/9711288].
%%CITATION = HEP-PH 9711288;%%

%\cite{Antusch:2005gp}
\bibitem{Antusch:2005gp}
S.~Antusch, J.~Kersten, M.~Lindner, M.~Ratz and M.~A.~Schmidt,
%``Running neutrino mass parameters in see-saw scenarios,''
JHEP {\bf 0503} (2005) 024
[arXiv:hep-ph/0501272].
%%CITATION = HEP-PH 0501272;%%

\end{thebibliography}
\end{document}